\documentclass[futureinternet,article,accept,moreauthors,pdftex,10pt,a4paper]{mdpi}
\usepackage{array}
\usepackage{rotating}
\usepackage{chngcntr}
\usepackage{epstopdf}

\usepackage{booktabs} 
\usepackage{multirow}
\usepackage{soul}
\usepackage{microtype}

\usepackage{booktabs} 
\usepackage{multirow}
\usepackage{soul}
\usepackage{microtype}
\newcommand\myurl[1]{\changeurlcolor{black}\url{#1}\changeurlcolor{blue}}
\makeatletter
\g@addto@macro{\UrlBreaks}{\UrlOrds}
\makeatother

\firstpage{1} 
\articlenumber{x}
\doinum{10.3390/------}
\pubvolume{8}
\pubyear{2016}
\copyrightyear{2016}
\externaleditor{Academic Editor: Emilio Ferrara}
\history{Received: 16 February 2016; Accepted: 29 June 2016; Published: 8 July 2016}

\Title{Conflict and Computation on Wikipedia: \mbox{A Finite-State} Machine Analysis of Editor Interactions}

\Author{Simon DeDeo$^{1,2,3,4}$} %Please carefully check the accuracy of names and affiliations. Changes will not be possible after proofreading. This is correct.
\AuthorNames{Simon DeDeo}
\address{%
$^{1}$ \quad Center for Complex Networks and Systems Research, Department of Informatics, Indiana University, \mbox{919 E 10th St}, Bloomington, IN 47408, USA; sdedeo@indiana.edu; Tel.: +1-812-856-2855; \mbox{Fax: +1-812-856-3825}\\
$^{2}$ \quad Program in Cognitive Science, Indiana University, 1900 E 10th St, Bloomington, IN 47406, USA\\
$^{3}$ \quad Ostrom Workshop in Political Theory and Policy Analysis, 513 N Park Avenue, Bloomington, IN 47408, USA\\ % the Ostrom Workshop does not have departments; this is the most specific form, and is standard
$^{4}$ \quad Santa Fe Institute, 1399 Hyde Park Road, Santa Fe, NM 87501, USA\\ % the Santa Fe Institute does not have departments; this is the most specific form, and is standard
}

\abstract{What is the boundary between a vigorous argument and a breakdown of relations? \mbox{What drives} a group of individuals across it? Taking Wikipedia as a test case, we use a hidden Markov model to approximate the computational structure and social grammar of more than \mbox{a decade} of cooperation and conflict among its editors. Across a wide range of pages, we discover a bursty war/peace structure where the systems can become trapped, sometimes for months, in a computational subspace associated with significantly higher levels of conflict-tracking ``revert'' actions. Distinct patterns of behavior characterize the lower-conflict subspace, including tit-for-tat reversion. While a fraction of the transitions between these subspaces are associated with top-down actions taken by administrators, the effects are weak. Surprisingly, we find no statistical signal that transitions are associated with the appearance of particularly anti-social users, and only weak association with significant news events outside the system. These findings are consistent with transitions being driven by decentralized processes with no clear locus of control. Models of belief revision in the presence of a common resource for information-sharing predict the existence of \mbox{two distinct} phases: a disordered high-conflict phase, and a frozen phase with spontaneously-broken symmetry. The bistability we observe empirically may be a consequence of editor turn-over, which drives the system to a critical point between them.}

\begin{document}

\section{Introduction} %we added the heading of the section, please confirm -- CONFIRMED

Many societies are characterized by periods of cooperation separated by significant periods of intra-group conflict. Both cooperation and conflict are themselves complex phenomena: cooperation is not the absence of conflict, nor is the presence of conflict incompatible with cooperation. One can talk of a lull in fighting, for example, or, conversely, of antagonistic interactions that are part of the internal logic of a resilient and cooperative society. Two questions, then suggest themselves: can we define the boundary between cooperation and conflict, and can transitions across this boundary be explained, controlled, or predicted before they happen?

\textls[-10]{We approach this pair of fundamental questions by examining a social system in terms of how it processes information. As we shall show, a system's computational process---how it moves through an abstract state space that both responds to past antagonism and predicts its future development---connects directly to large-scale features that are immediately perceptible to human participants.} 

Our system of study is social behavior observed in the debates on the website Wikipedia. Wikipedia is an ideal case study for the computational properties of conflict. By design, actions on the system are logged at extremely high resolution, allowing us to track the actions and interactions of tens of thousands of users second-by-second for nearly fifteen years. A natural binary classification of actions (``revert'' and ``non-revert''), that tracks conflict between users, allows us to use sequence analysis and the language of symbolic timeseries~\cite{dedeo2013collective,keegan2015analyzing} to operationalize the boundary between conflict and cooperation with Hidden Markov Models (HMMs).

Using HMMs to detect hidden structure in social behavior is a natural extension to how they have been used in the engineering, linguistic, and biological sciences. HMMs are in wide use when we believe that a noisy observable signal tracks some more complex underlying process, such as when the garbled sound of a noisy room includes someone speaking a sentence (speech recognition~\cite{speech1,speech2}), the individual words in a sentence give clues to that sentence's hidden syntactic structure (part of speech tagging~\cite{derose1988grammatical,church88}), or the A, C, T, and G symbols of DNA code for genes~\cite{Salzberg01011998}. The volume of data on Wikipedia makes it possible to model the information processing of the system itself, an approach becoming increasingly common in large-scale studies of social media~\cite{darmon2013predictability}. At the same time, our understanding of the products of this analysis are enhanced by ethnographic and quantitative studies of the system~\cite{reagle2010good} that allow us to interpret individual actions, users, and their social contexts, and allow us to test narrative accounts of the causes and patterns of conflict.

Our analysis of Wikipedian conflict through the lens of the revert/non-revert distinction reveals two long-lived behavioral patterns associated with higher and lower levels of these conflict-tracking revert actions. While the modes themselves are defined by reference to short sequences, or motifs, of interactions that can take only a few minutes to happen, the system as a whole will stay in one mode or another for weeks and, sometimes, even months or years. This separation of timescales provides a bottom-up definition of the boundary between cooperation and conflict that depends on the existence of persistent, hidden system memories. In the first part of this paper, we demonstrate the existence of these unexpectedly long-term memories, and the distinct local patterns of interaction to which they give rise. Detected by reference to apparently abstract properties of the system's computational description using hidden Markov models, these memories are tied to human-scale and group-level properties that form the basis of the system's self-understanding.

In addition to providing a novel means for the description of the boundaries and distinctions between conflict and cooperation, our model allows us to pin-point the transitions between these \mbox{two coarse-grained} computational states. In the second part of our paper, we use this information to study the influence of administrative actions, individual users, and exogenous events over this society's history. We find that top-down actions have only limited effectiveness in triggering a transition in either direction.

\section{Methods}

Our study here is based on the time series of edits made on sixty-two of the most-edited pages on Wikipedia. This sample was chosen because each page then provided us with sufficient data to model each page independently; we thus were able to avoid making the (strong) assumption that editing practices were uniform. While our sample is by definition exceptional, and includes articles with tens of thousands of individual edits, it also covers a wide range of different topics and themes including (1) biographies of figures in both politics and entertainment, living and dead, from Genghis Khan to George W. Bush, Paul McCartney to Britney Spears (30\% of the sample); (2) other non-biographical pages associated with the arts and entertainment (16\% of the sample); (3) pages associated with countries, such as Cuba, Argentina and the United States (18\% of the sample); (4) natural events ranging from the sinking of the Titanic to Hurricane Katrina; (5) political events, such as World War II and the 2006 War in Lebanon (13\% of the sample); (6) technology topics, such as the iPhone (9\% of the sample); and (7) religious topics, such as the Catholic Church and Islam (6\% of the sample). See Appendix~\ref{full_list}; past work on Wikipedia editing practices has ranged from the histories of individual articles~\cite{kittur2007he, kane2014} to the full Wikipedia dataset~\cite{kriplean2007}; ours is of a similar scale and selectivity to a number of other conflict studies~\cite{viegas2004studying,kane2011,arazy2011}, and the article list overlaps with two prior studies~\cite{dedeo2013collective,group}.

Data for this paper, and accompanying analysis code, can be found at~\cite{open_data} and~\cite{sfihmm}. % SFIHMM is the name of a computer program; it is now referenced

\subsection{Tracking Conflict through Page Reverts}

%% this paragraph defines R (a revert) and C (a non-revert).
When editors interact on Wikipedia, they can disagree on what to do and how to behave. \mbox{The resulting} conflicts that emerge can be tracked by reference to ``reverts''. Reverts are when one user takes a page to a previously-seen state, effectively undoing and discarding the work of others. \mbox{For each} page in our sample, the edit time series was coarse-grained into ``revert'' (R) and non-revert (C) actions, as described in Ref.~\cite{dedeo2013collective}. We define a revert as an edit that takes the page to a previously-seen state (as detected by the MD5 page hash provided by the Wikipedia API), but exclude the relatively small number of cases where a user's revert only undoes work contributed by that user herself (i.e., we exclude self-reverts). More details on these methods, and additional robustness checks, are described in~\cite{group}.

Reverts are noisy, imperfect signals of conflict; research has shown how more qualitative, interpretive, ``thicker''~\cite{geertz1994thick} measures of conflict among users are strongly associated with these easily-tracked reverts~\cite{yasseri2012dynamics}, which can degenerate into edit wars~\cite{viegas2004studying}. Researchers have shown that reverts play a central role in marking both conflict and controversy~\cite{kittur2009,ref3,ref35}, and that a great deal of information is encoded in reverts, including patterns of alliances among editors~\cite{kittur2007he}. \mbox{Reverts play} a major role in the social norms of Wikipedia itself~\cite{bradi}, including highly central pages in the norm network such as the ``three-revert rule'', and strong norms against edit warring (see~\cite{group} for \mbox{further discussion}).

Reverts, however, are not equivalent to conflict. In general, reverts are best used to track the presence of task conflict~\cite{arazy2011}, as opposed to other dimensions of conflict identified in the organizational behavior literature, such as affective conflict and process conflict~\cite{hb,jehn}. While the task conflict represented by reverts is more ambiguous than other forms of conflict, it is still highly correlated with affective and relational conflict~\cite{kitturkraut2010}; for example, the perception of rejection associated with being reverted is a significant factor in dissuading female contributors from participation~\cite{collierbear2012}. At the same time, the tension between the need to change and to retain text can manifest itself in many different and more subtle ways~\cite{auray2007democratizing,joe,kane2014}. There are multiple pathways for conflict to manifest itself that do not lead deterministically to reverts, an example being conflict associated with Wikipedia policies~\cite{kriplean2007} and conflict that takes place in discussion on article talk pages~\cite{viegas2007}. 

In short: while we rely on reverts in our analysis here, we emphasize that this is only one signal of conflict on Wikipedia and that conflict itself is a far richer notion than the binary presence-absence data of our time series. The fact that reverts are only noisy, partial signals of conflict leads us to use \mbox{a particular} tool, the hidden Markov model. Hidden Markov models are ideally suited to cases where a complex system leaves only a low-dimensional trace in an observable time-stream, allowing us to reconstruct, in part, some of the structure of the hidden, underlying process.

\subsection{Hidden Markov Models}

For each page, we have a binary time series of Wikipedian conflict, consisting of Rs and Cs. Page-by-page, we approximate this time series as a probabilistic finite state machine, more commonly known as a Hidden Markov Model (HMM, or machine). In an HMM, the observed behavior of the system---here the registering of a new edit of type R or C---is (probabilistically) conditioned on the system's position in a hidden, internal state space. 

\begin{figure}[H]
\centering
\begin{tabular}{c}
\includegraphics[width=0.75\textwidth]{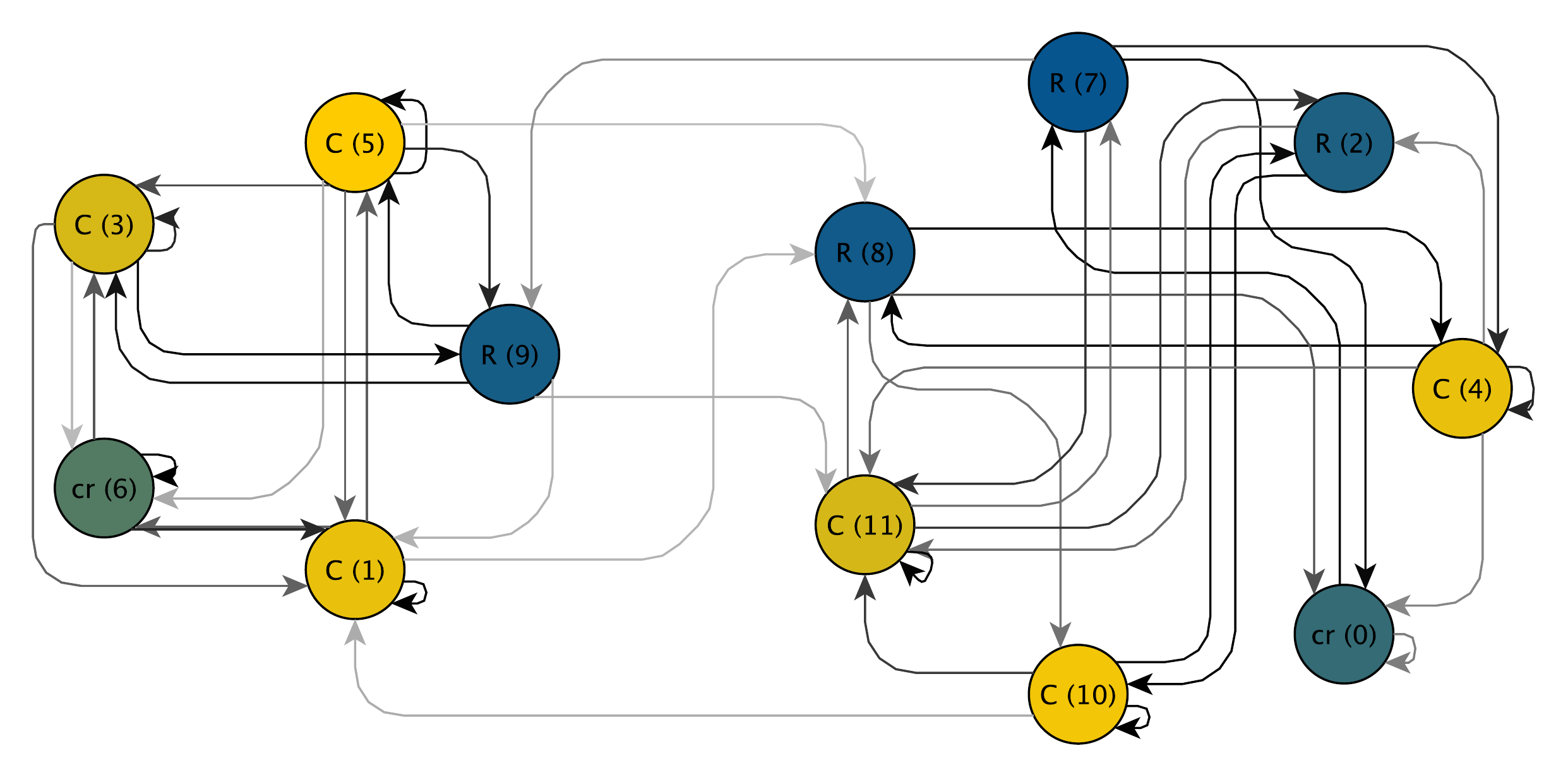} \\ \hline
\begin{tabular}{lc|c}
Page Behavior & {\tt ...CRCRRRCCCCCCRRRCCCCC}& {\tt RCRCRCRCRCRCRCRCRRRRR...} \\
Internal State & {\tt ...59393955555593955555} & {\tt 848484848484848480707...} \\ 
\end{tabular} \\ \hline
\end{tabular}
%%% C and R already defined. cr is not an acronym, just a label, and should not be capitalized. UTC is now defined.
\caption{{Top panel}: Hidden Markov model for cooperation and conflict on the George W. Bush page. States are labeled by the probability of emitting each of the two output symbols; ``C'' ($P(C)> 0.8$, lighter yellow), ``R'' ($P(R)> 0.8$, darker blue) or ``cr'' (otherwise). Edge weights show transition probabilities; the lightest lines, connecting the two subspaces, correspond to probabilities of order $10^{-4}$. Despite its complexity, the system is dominated by a transition logic that, on short timescales, confines the system to one of two separate modules with a high density of internal transitions. {Bottom panel}: below the diagram, as an example of our method, we show the Viterbi reconstruction in the neighborhood of one of the sixteen between-subspace transitions in our data. As the system wanders, during this period, among internal States 5, 9, 3, 8 and 4, it emits the symbols C and R probabilistically depending on which particular state it is in. On 8 November 2004 at 9:11 UTC (Coordinated Universal Time), long runs of cooperation and conflict gave way to more rapid-fire conflict and vandal-repair as the system crossed from the left module to the right, via the bridge between State 5 and State 8.\label{gwb_picture}} 
\end{figure}

Our use of the HMM is conceptually simple. Each page is associated with a distinct machine. \mbox{The evolution} of the page is represented by the machine as it moves from one hidden, internal state to another, from edit to edit. At each edit, and depending on the state that the machine finds itself in, it will produce an observed edit behavior (probabilistically; here, R or C), and transition to a new internal state. The chance of going from one hidden state to another is dictated by a fixed probability. \mbox{The origins} of the HMM paradigm go back to~\cite{baum1966}; for a recent, technical introduction to hidden Markov models, including the Baum--Welch algorithm used in this paper, see~\cite{press2007numerical}; Figure~\ref{gwb_picture} provides an example from the George W. Bush page. %%FIGURE GWBALT HERE

In contrast to many simple time series models, a hidden Markov model does not make a ``no history'' assumption: instead, the system memory is encoded by the position within the hidden state space. From the point of view of an external observer, this memory can extend arbitrarily far into the past. Strong theorems exist about the way in which this memory must, eventually, decay, but there is no ``hard'' cutoff where (for example) an observed event $t$ steps in the past has zero influence when $t$ is greater than some critical, finite $T$ and one includes the more recent past as a predictor. For these reasons, an HMM is more general than the standard auto-regressive (AR~\cite{mills1991time}) models familiar from econometrics. AR models condition the current state of the system on a fixed and finite number of steps of the observable sequence in the recent past; usually less than ten. By contrast, an HMM can encode arbitrarily long path dependencies; as we shall see, the correlation lengths our methods do detect in Wikipedia suggest that AR models would be very poor descriptions of the data indeed.

\subsection{Fitting and Characterizing HMMs}

We find the parameters for the HMM associated with the time series on each page using the Expectation-Maximization algorithm (EM~\cite{baum1966,press2007numerical}); for each page, we run this hill-climbing algorithm 3200 times with random initial seeds to study the ruggedness of the underlying likelihood landscape. 

The one externally-fixed parameter in an HMM is the total number of hidden states; once this choice is made, the other parameters are determined by the data. Increasing the number of states means that the model is better able to predict the data. Using too many states, however, can mean that the model structure includes features that are due to random, one-off fluctuations in the data; this is called over-fitting. We use the Akaike Information Criterion (AIC~\cite{akaike1974new}) to select the number of states for each page's model. Extensive tests, and a comparison to the main competing model selection method, the Bayesian Information Criterion (BIC~\cite{schwarz1978estimating}), suggest that AIC is, for the systems we encounter here, preferred; see Appendix~\ref{model_selection} for further discussion. For simplicity, and in keeping with usual practice, we work solely with the maximum-likelihood model, rather than attempting to model the full distribution over distinct, but less-preferred, models.

\subsection{Subspaces, Trapping Time, and Viterbi Reconstruction for HMMs}

In general, as a system evolves in the HMM paradigm, it will spread out over its internal state space. A characteristic feature of this spreading out is the so-called relaxation time, which determines how long the system stays confined in a particular subspace of its possible internal states before returning to its characteristic stationary distribution. Once the best-fit HMM has been found, we can compute the relaxation time, $\tau$, by reference to the second eigenvalue of the transition matrix, $\lambda_2$,
\begin{equation}
\tau = \frac{1}{1-\lambda_2}.
\label{relax}
\end{equation}

Informally, and for the cases we consider here, when $\lambda_2$ is very close to unity (small spectral gap), the system can take a long time to return to the stationary distribution (see Appendix~\ref{formal_lambda2} for further discussion). This can be understood by writing down an approximation for the probability that the system will be found in a particular state at time $t$, $\vec{v}(t)$,
\begin{eqnarray}
\vec{v}(t) & \propto & v^{(1)} + \alpha v^{(2)}\lambda_2^t + \ldots, \nonumber \\
& \approx & v^{(1)} + \alpha v^{(2)} e^{-t/(\tau-\frac{1}{2})} + \ldots \label{decay}
\end{eqnarray}
where $v^{(1)}$ is the stationary distribution of the chain, equal to the first eigenvector of the transition matrix, $v^{(2)}$ is the second eigenvector, and $\alpha$ depends on the initial conditions. The remaining terms are exponentially suppressed relative to $v^{(2)}$; the approximation in the second line holds in the limit that $\lambda_2$ is close to unity.

The second eigenvector, $v^{(2)}$, describes the perturbation that takes the longest time to die away. It allows us to split the system into two subspaces based on the signs of its entries; we define all of the states where $v^{(2)}$ is, say, positive, as ``subspace one'', and the remaining states as ``subspace two''. One set of states is associated with the positive $v^{(2)}$ values; the complementary set with those that are strictly negative. Informally, initial conditions that are weighted towards states solely in one of these two sets (all positive $v^{(2)}$ values or all negative) take the longest to decay to the stationary distribution. It is these two spaces that will define the system epochs. 

The use of the sign structure of an HMM's eigenvectors is a natural way to identify these subspaces; similar methods have been applied to, for example, the identification of metastable states in chemical reaction networks through what is called Perron Cluster Cluster Analysis (PCCA)~\cite{deuflhard1999computation,cordes2002metastable,deuflhard2005robust}. Use of just the second eigenvector naturally splits the system into the two most significant subspaces. \mbox{The sign} structure of the third and higher eigenvectors can be used to further decompose the space in a hierarchical fashion, potentially subdividing these larger modules (see, e.g.,~\cite{noe2007hierarchical}). In this paper, we focus solely on the first division, and do not attempt to identify substructures within either of the two main clusters.

Once we know the parameters of the HMM, we can use Viterbi path reconstruction~\cite{forney1973viterbi} to reconstruct the maximum-likelihood path through the state space. We then know, at any point in the time series, where we are in the underlying computational state space. 

This allows us to associate an internal state of the HMM to each step in the time series and, thus, to each edit on the page itself. We can also pinpoint when a transition from one subspace to the other occurs. When a system switches subspaces, there can be some flickering, with rapid shifting back and forth between subspaces before the system settles down. We only count a transition from one subspace to another when the system remains in the new subspace for more than ten time steps.

Given this definition, we can then study time spent within each subspace; we call this empirical quantity the ``trapping time'', $\tilde{\tau}$; per the definition of $\tau$ and the Levin--Peres--Wilmer theorem (see Appendix~\ref{formal_lambda2}), we expect the trapping time to be of order $\tau$. We report both the trapping time and the relaxation time in our results here. The trapping time is the more empirical quantity, since it describes the actual behavior of the system: what state the system was actually in at a particular time, and how long the system was trapped in one subspace or the other during its evolution; we can report $\tilde{\tau}$ for each subspace separately or just consider the average. Meanwhile, the relaxation time is the more theoretical quantity, characterizing the HMM in isolation, in terms of the generic properties of the kinds of time series it tends to generate. 

In order to test the extent to which $\tau$ (and $\tilde{\tau}$) are driven by system-wide patterns of linking between internal states (rather than just, for example, generic sparseness, or an overall tendency for any state to link to only a small number of other states), we consider a null model for $\tau$, where we shuffle the entries of each state's probability vector. This scrambles the overall structure machine while keeping the list of transition probabilities for each state constant.

On the simplest level, subspaces can be defined by their average levels of conflict: the fraction of time they lead to the emission of a revert symbol, R. We track this using the revert ratio: the fraction of reverts in the higher-conflict subspace divided by the fraction of reverts in the lower-conflict subspace. 

Subspaces are defined by more than just their levels of reverting. What distinguishes them in \mbox{a deeper} sense are the relationships between their hidden internal states, which make some sequences more likely than others. System response to a revert will be different, and lead to different characteristic futures, in one subspace compared to the other. 

%% partial-KL now defined
We characterize subspaces by reference to the relative frequency of motifs: short sequences of system behaviors. We measure the probability of different motifs in each subspace, and then consider the motifs with the highest partial-KL. In particular, if the probability of motif $i$ in the first subspace is $p_i$, and the probability of $i$ in the second subspace is $q_i$, we first define the mixture distribution, $m_i$, equal to $(1/2)(p_i+m_i)$. Then, the partial-KL for motif $i$ in the first subspace is $p_i\log{p_i/m_i}$. Partial-KL provides an information theoretic measure of the extent to which a particular motif is a signal of the underlying subspace~\cite{entropy}; a motif $i$ is characteristic of a subspace if the partial-KL is large; see Ref.~\cite{obo} for further discussion.

\subsection{Causes of State Transitions}

Given an identification of transition points, we return to the original time series to catalog features of the system at these critical points, and to determine potential causes for the transition. We consider three types of potential causes. In all three cases, we look for associations between these potential causes, and transitions between subspaces. An association is defined as a potential cause occurring within a particular time window of a transition event.

First, we track page protection events, points at which administrators changed the access permissions of a particular page to either prevent or allow editing by different user classes. Records of page protection events are reliably logged beginning on 10 November 2003. Page protection is a crucial, hidden element of the system's conflict management, and has a significant impact on the composition of a page's editing population~\cite{hill2015page}. We expect page protection events to be the main source of top-down control that can switch the system between trapping subspaces. We characterize a protection event as ``hard'' when it leads to a restriction on who may edit the page; ``soft'' when it releases the page from \mbox{a restriction}. The bulk of protection events shift pages between ``anyone can edit'' and ``semi-protected''; the latter restricts edits to users who have been registered for a sufficient number of days, and have made a sufficient number of edits on other pages. 

Second, we track anti-social user events. Transitions between subspaces may, potentially, be induced by unusually anti-social users who instigate self-sustaining conflict by, for example, publicly violating the norms of interaction. Rather than define an externally-imposed standard that may not reflect the reality of online interaction, we define anti-sociality by reference to community norms, as expressed by so-called user blocking events. Wikipedia administrators are able to block users from editing for \mbox{a period} of time. Receiving a block is a signal that one has (in the opinion of at least one administrator) violated a community norm.

We define a user as anti-social in the context of a particular page if that user has a blocking rate (number of blocks per total number of edits) higher than 95\% of randomly-selected editors on the same page. For any transition, we check to see if the user with the most edits within a certain window meets this anti-social criterion; if so, we consider this a potential explanation of the transition. This provides a quantitative measure of the extent of internal opposition to dominant norms.

Finally, we track major external events; transition dates associated with major events concerning the topic of the page itself, and defined as significant increases in news coverage for the article subject.

To do this, we rely on two well-curated, public databases from the New York Times~\cite{nyt_api} the Guardian~\cite{guardian_api}, and track sudden increases in the density of news coverage. We quantify news spikes by reference to the rate of articles in a four-day window around the position in question, and define \mbox{a major} external event as one that leads to a fluctuation in this ratio over and above the null rate at 95\% confidence. We search for news articles using the full text of the article's title. %% references fixed

Time series data, resultant best-fit HMMs, and code for both estimation of Markov chains and for Viterbi reconstruction, are available online as the package SFIHMM~\cite{sfihmm}, and in the open data release~\cite{open_data}. % SFIHMM is the name of a computer program; it is now referenced

\section{Results}

For all sixty-two pages, the EM algorithm converged, and we were able to find best-fit hidden Markov models. For the majority of pages (56 of 62), model selection preferred models with at least six~internal states.

The network of transitions between states was sparse; any internal state had significantly non-zero transition probabilities to only a small number of others. States themselves usually had near-deterministic emission rules; a particular state usually had a near-unity probability of emitting \mbox{one of} the two possible R or C symbols. Informally, this implies the existence of a variety of interaction motifs, particular patterns of Rs and Cs, at the multi-symbol level.

An example of the underlying computation process for a page is shown in Figure~\ref{gwb_picture}, for the most-edited page in Wikipedia: that associated with George W. Bush. The sparseness of the internal state connectivity can be seen in the small number of high-probability transitions for each node; the determinism of the internal states from the fact that nearly all states are strongly biased, at at least 80--20, towards one or other of the two symbols. Also clearly visible is the modular structure of the hidden system; most transitions occur within one of two clearly separated subsets of the states. \mbox{Over the} course of nearly fifteen years and 45,448 edits, the page switched between these two subspaces only sixteen times; we show the time series surrounding one switch, which happened on 8 November 2004. The motifs on either side of the transition are visibly distinct; we discuss and quantify this effect below.

Table~\ref{topten} then shows the recovered machines for the top ten pages by number of edits; the layout for each of the machines is dictated by the recovered modules, with states that are in the same module grouped together visually. 
\newpage
% change it to landscape
\paperwidth=\pdfpageheight
\paperheight=\pdfpagewidth
\pdfpageheight=\paperheight
\pdfpagewidth=\paperwidth
\newgeometry{layoutwidth=297mm,layoutheight=210 mm, left=2.7cm,right=2.7cm,top=1.8cm,bottom=1.5cm, includehead,includefoot}
\fancyheadoffset[LO,RE]{0cm}
\fancyheadoffset[RO,LE]{0cm}
%%%%%%%%%%%%%%%%%%%%%%%%%%%%%%%%%%%%%%%%%%%%%%%
\begin{table}[H]
\centering
\begin{tabular}{m{1.2in} m{2in} | m{1.2in} m{2in}}
{\tiny \begin{tabular}{l} {\tt George\_W.\_Bush } \\ $\tau$: 1451 steps \\ High/Low Conflict Ratio: 1.83 \\ Median edit-time (minutes): \\ High: 3.53; Low: 15.85 \\ Average trapping time (months): \\ High: 3.71; Low: 13.63 \end{tabular}} & \includegraphics[width=1.6in]{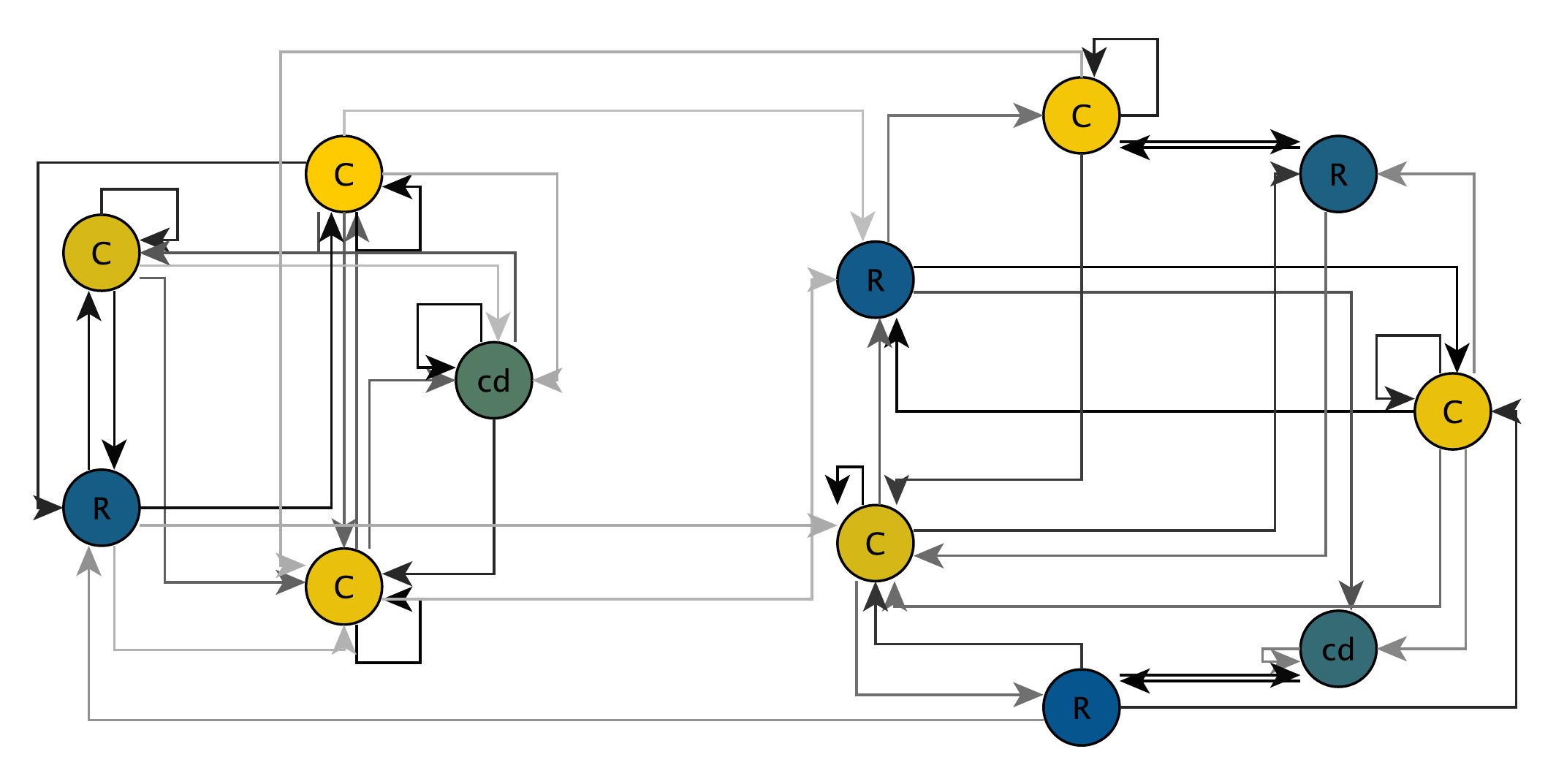} & {\tiny \begin{tabular}{l} {\tt United\_States } \\ $\tau$: 360 steps \\ High/Low Conflict Ratio: 3.61 \\ Median edit-time (minutes): \\ High: 13.48; Low: 14.78 \\ Average trapping time (months): \\ High: 5.78; Low: 4.2 \end{tabular}} & \includegraphics[width=1.6in]{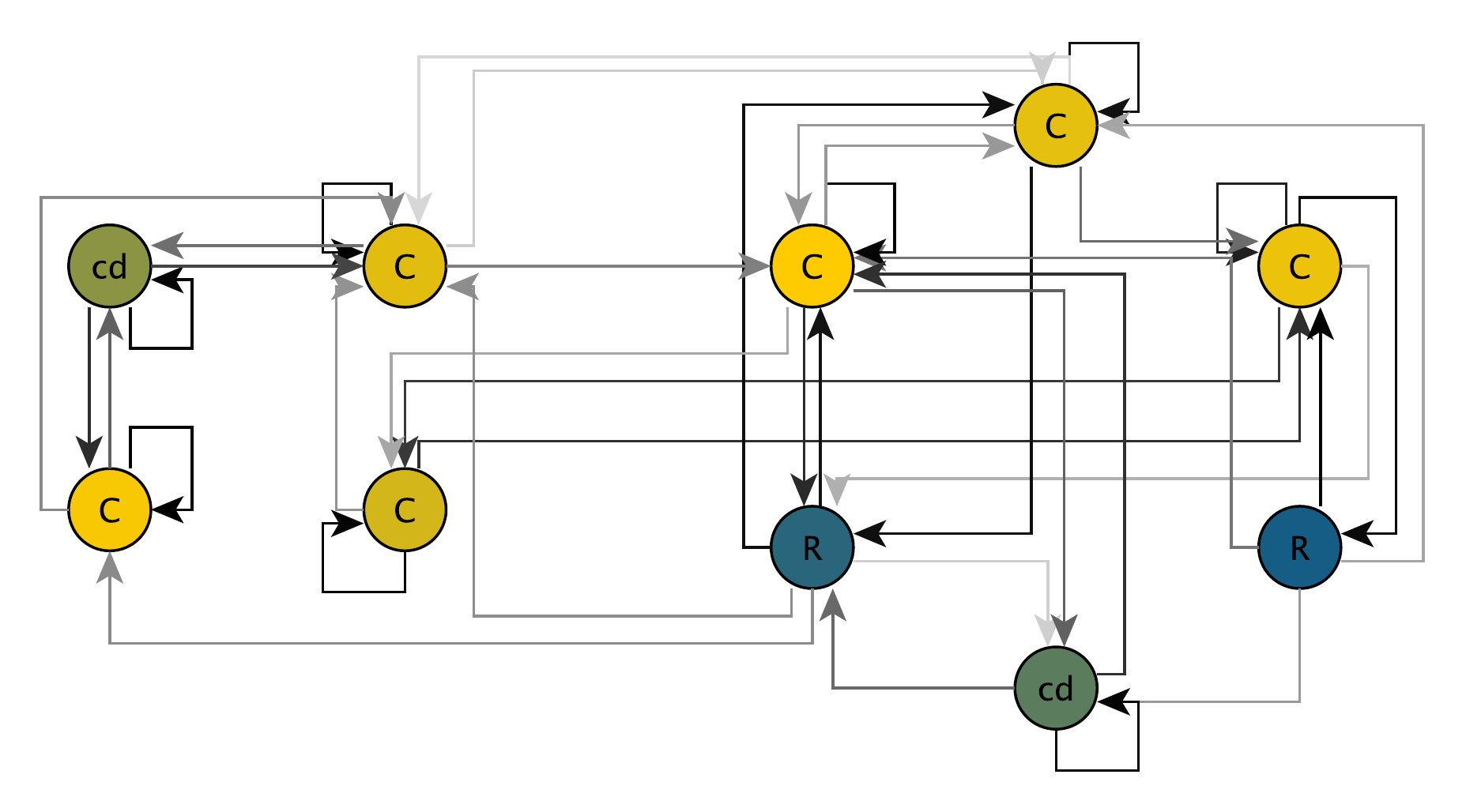} \\ 
{\tiny \begin{tabular}{l} {\tt United\_States } \\ $\tau$: 360 steps \\ High/Low Conflict Ratio: 3.61 \\ Median edit-time (minutes): \\ High: 13.48; Low: 14.78 \\ Average trapping time (months): \\ High: 5.78; Low: 4.2 \end{tabular}} & \includegraphics[width=1.6in]{United_States.pdf} & {\tiny \begin{tabular}{l} {\tt Michael\_Jackson } \\ $\tau$: 227 steps \\ High/Low Conflict Ratio: 2.09 \\ Median edit-time (minutes): \\ High: 7.75; Low: 8.42 \\ Average trapping time (months): \\ High: 0.99; Low: 7.63 \end{tabular}} & \includegraphics[width=1.6in]{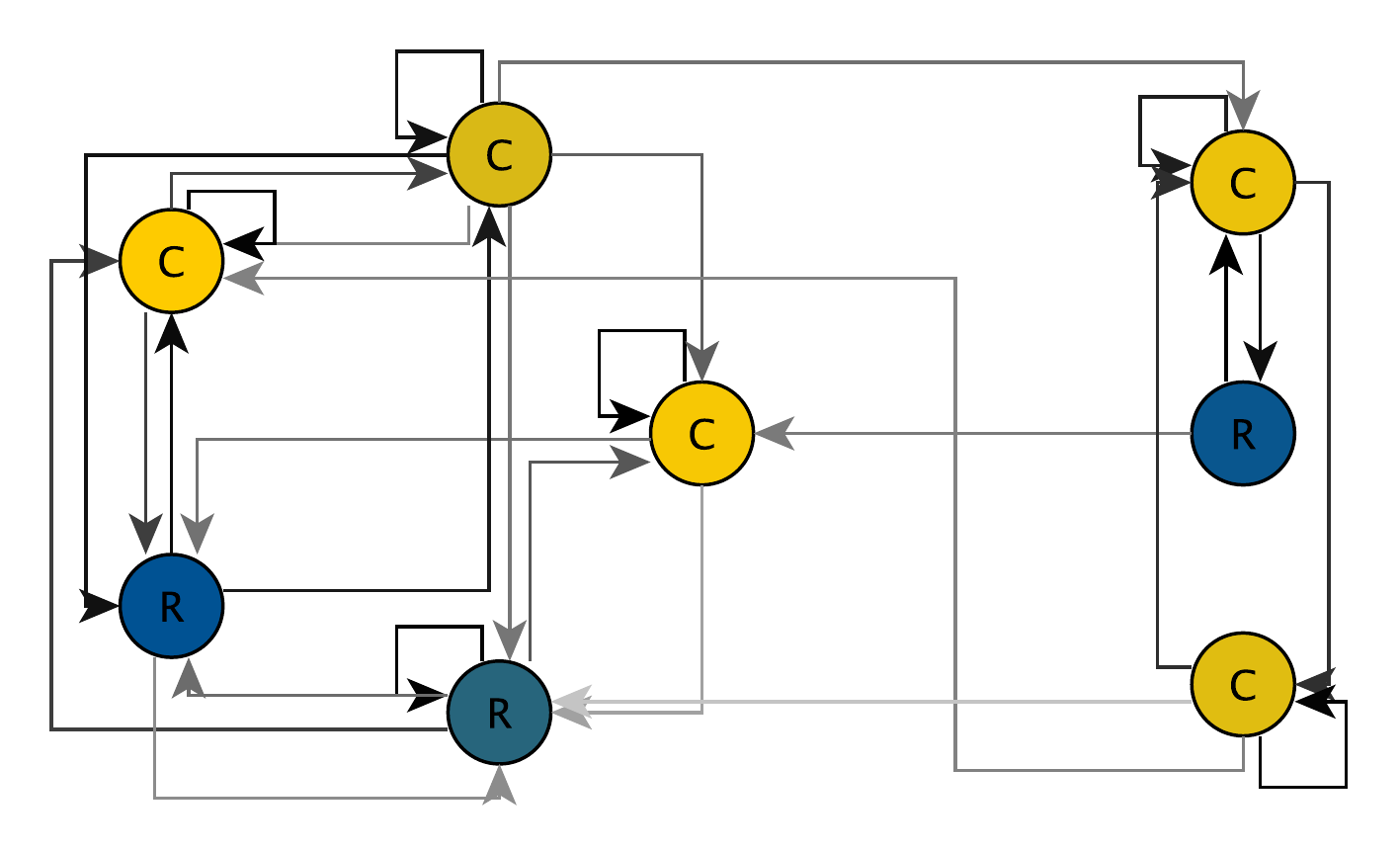} \\ 
{\tiny \begin{tabular}{l} {\tt Wikipedia } \\ $\tau$: 3695 steps \\ High/Low Conflict Ratio: 1.75 \\ Median edit-time (minutes): \\ High: 2.58; Low: 19.6 \\ Average trapping time (months): \\ High: 7.28; Low: 45.8 \end{tabular}} & \includegraphics[width=1.6in]{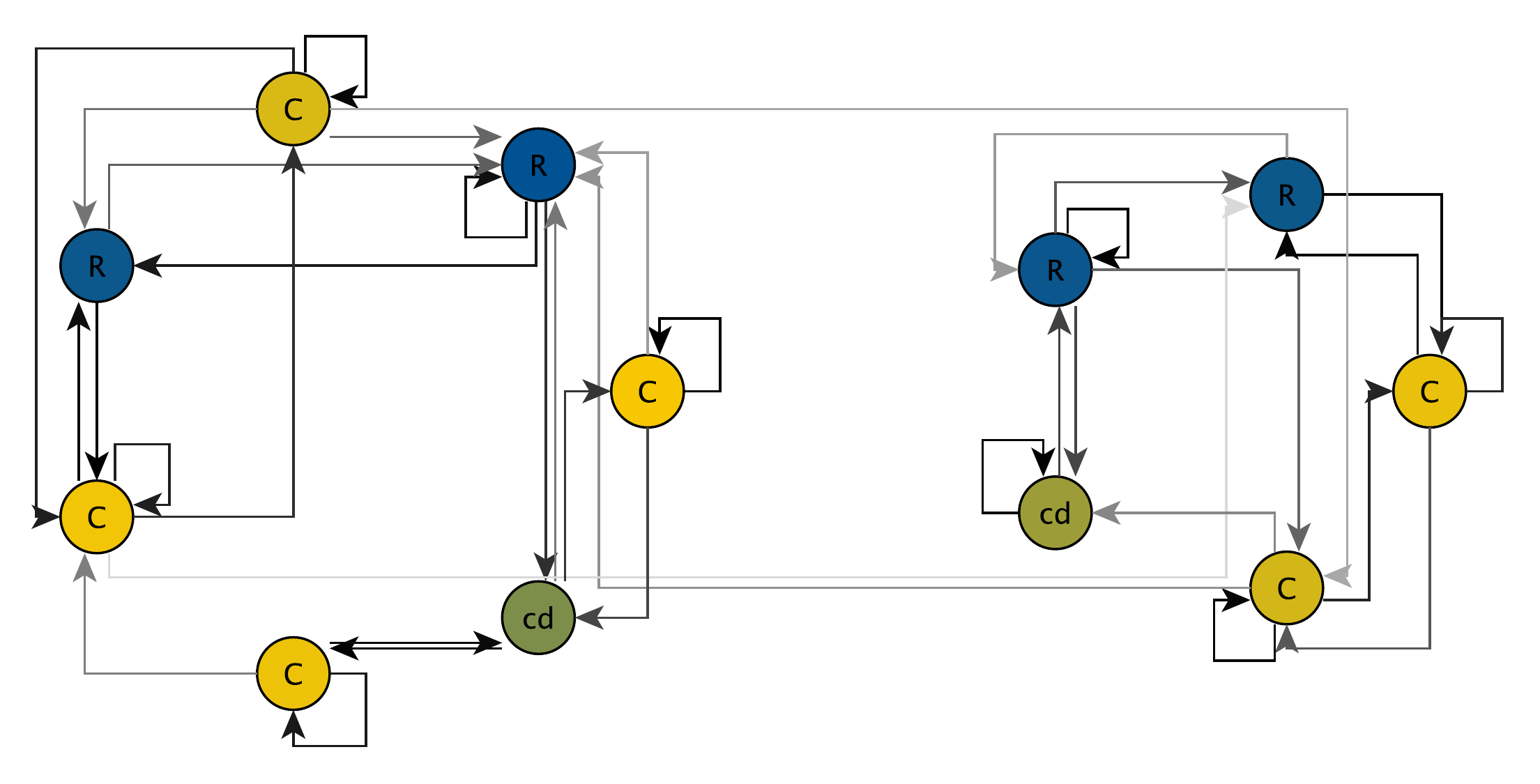} & {\tiny \begin{tabular}{l} {\tt Barack\_Obama } \\ $\tau$: 53 steps \\ High/Low Conflict Ratio: 2.31 \\ Median edit-time (minutes): \\ High: 17.58; Low: 7.73 \\ Average trapping time (months): \\ High: 1.82; Low: 0.46 \end{tabular}} & \includegraphics[width=1.6in]{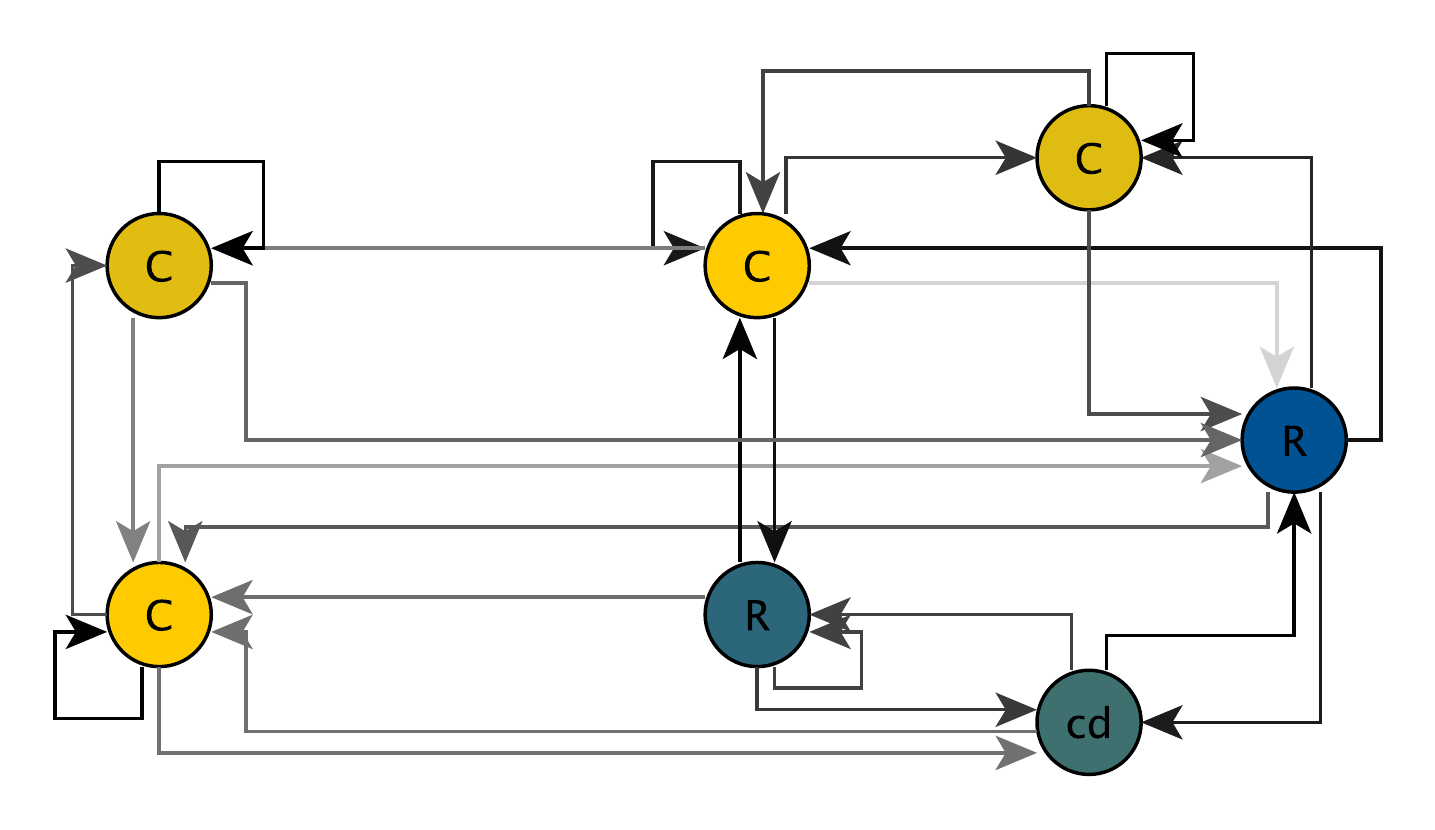} \\ 
{\tiny \begin{tabular}{l} {\tt Michael\_Jackson } \\ $\tau$: 227 steps \\ High/Low Conflict Ratio: 2.09 \\ Median edit-time (minutes): \\ High: 7.75; Low: 8.42 \\ Average trapping time (months): \\ High: 0.99; Low: 7.63 \end{tabular}} & \includegraphics[width=1.6in]{Michael_Jackson.pdf} & {\tiny \begin{tabular}{l} {\tt Global\_warming } \\ $\tau$: 277 steps \\ High/Low Conflict Ratio: 2.23 \\ Median edit-time (minutes): \\ High: 9.72; Low: 19.45 \\ Average trapping time (months): \\ High: 2.27; Low: 7.55 \end{tabular}} & \includegraphics[width=1.6in]{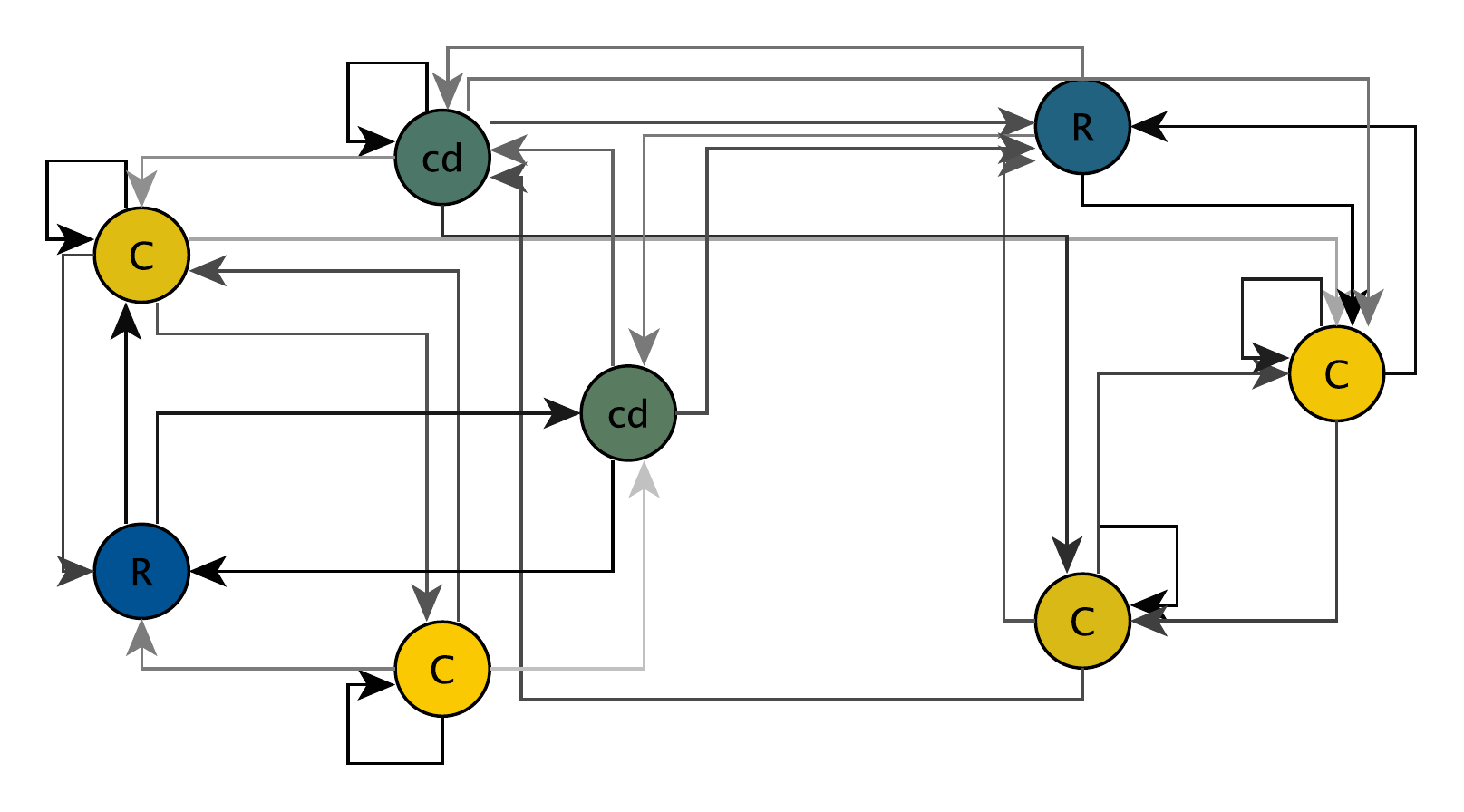} \\ 
{\tiny \begin{tabular}{l} {\tt Catholic\_Church } \\ $\tau$: 2357 steps \\ High/Low Conflict Ratio: 18.09 \\ Median edit-time (minutes): \\ High: 13.72; Low: 4.63 \\ Average trapping time (months): \\ High: 61.45; Low: 2.55 \end{tabular}} & \includegraphics[width=1.6in]{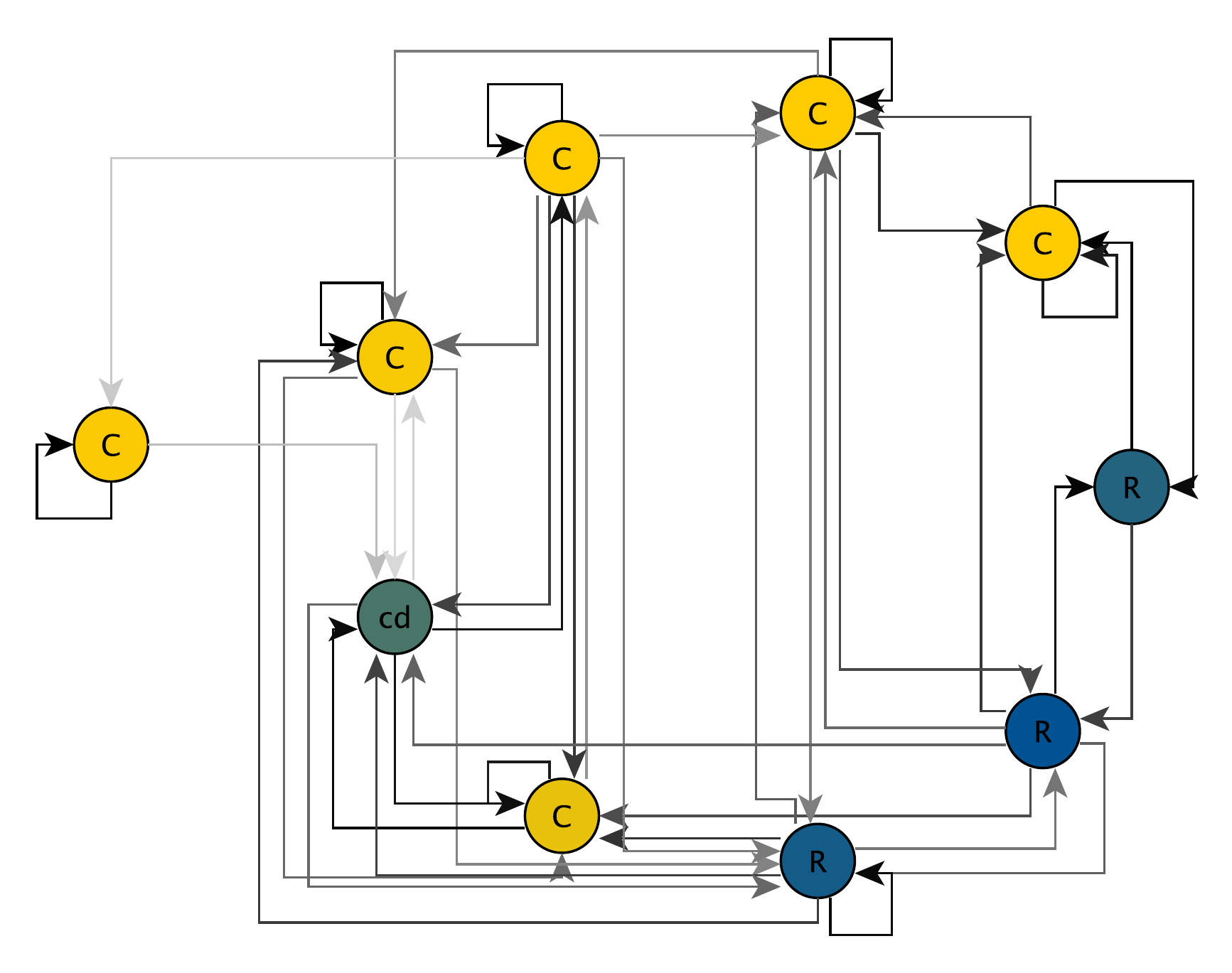} & {\tiny \begin{tabular}{l} {\tt Islam } \\ $\tau$: 287 steps \\ High/Low Conflict Ratio: 2.67 \\ Median edit-time (minutes): \\ High: 12.02; Low: 18.13 \\ Average trapping time (months): \\ High: 3.16; Low: 7.87 \end{tabular}} & \includegraphics[width=1.6in]{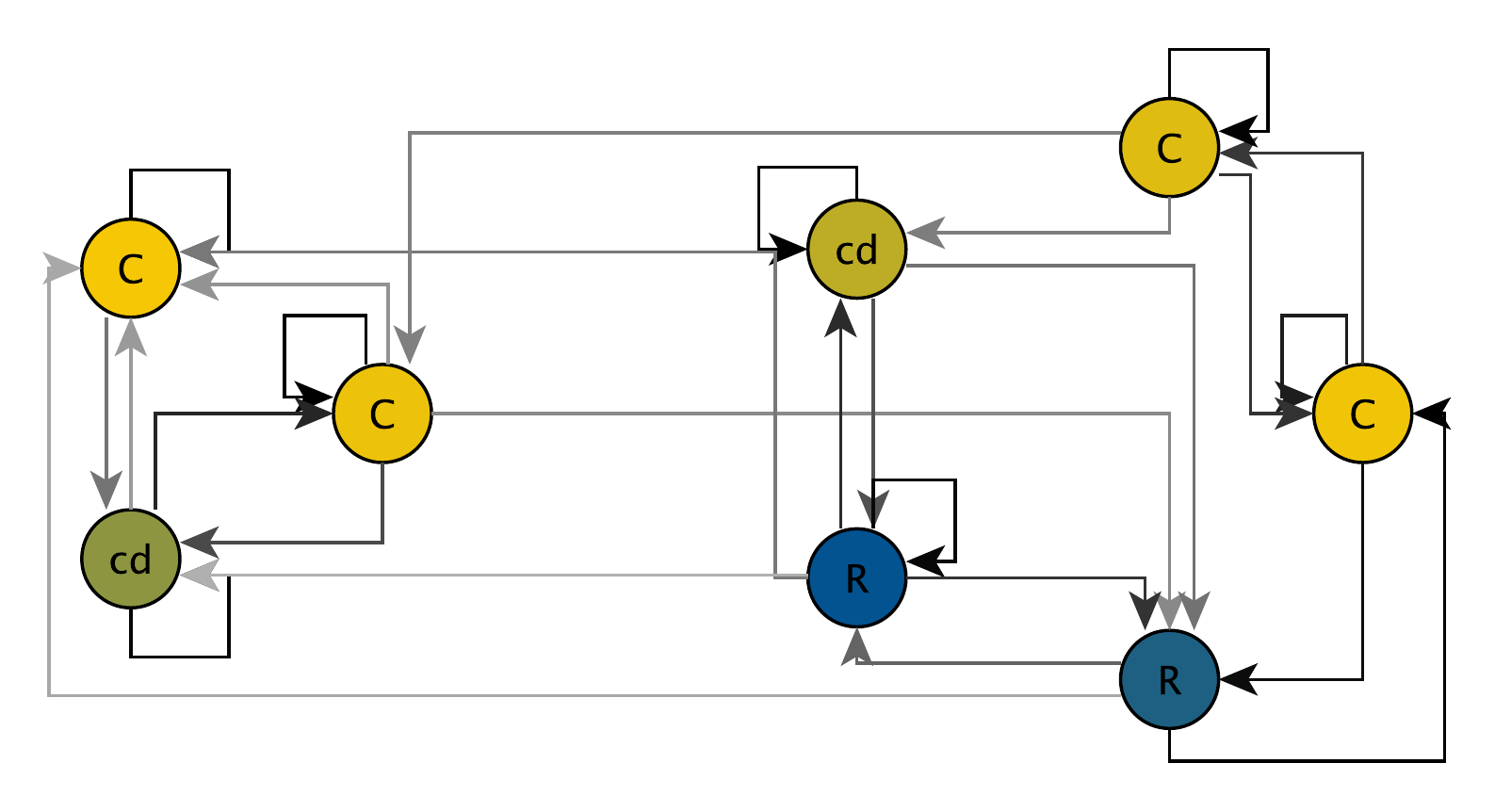} \\ 
\end{tabular}
\caption{Hidden Markov models and derived parameters for cooperation and conflict on the ten most-edited pages on Wikipedia. Editing patterns are characterized by high levels of determinism, and long timescale trapping in distinct higher and lower conflict spaces.\label{topten}}
\end{table}
%%%%%%%%%%%%%%%%%%%%%%%%%%%%%%%%%%%%%%%%%%%%%%%
% change everything back
\newpage
\restoregeometry
\paperwidth=\pdfpageheight
\paperheight=\pdfpagewidth
\pdfpageheight=\paperheight
\pdfpagewidth=\paperwidth
\headwidth=\textwidth

\subsection{Epoch Detection}

\begin{figure}[H]
\centering
\includegraphics[width=0.75\textwidth]{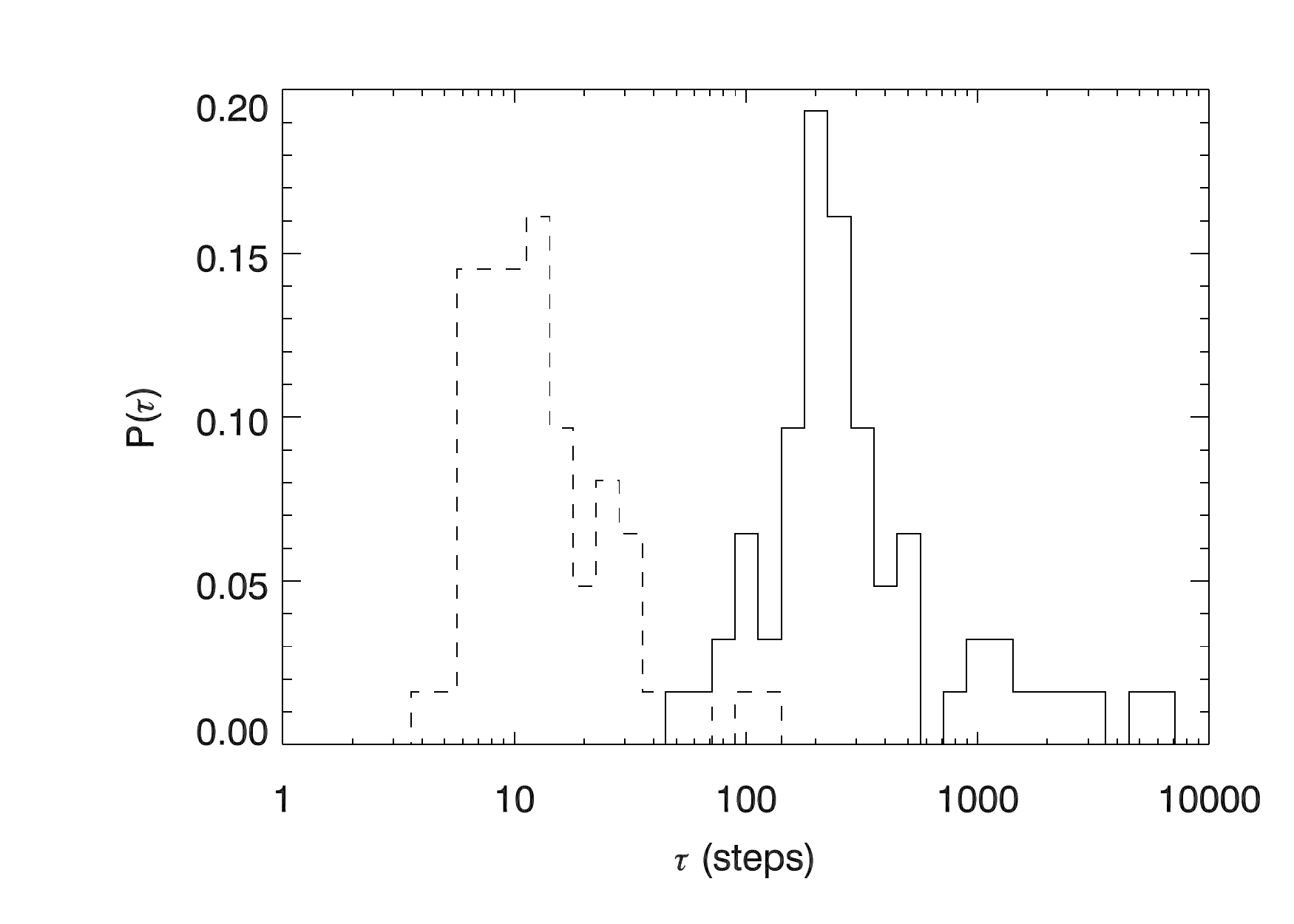}
%% font size fixed; thanks
\caption{Relaxation time, $\tau$, for the sixty-two pages in our sample (solid line). Times are exceptionally long, an average of $698$ steps, and, on average, a factor of $50$ times longer than expected for Markov models with similar sparseness (dotted line). The longest trapping times are for the pages associated with the Gaza War, the Russo-Georgian War, and the page describing Wikipedia itself.\label{tau}}
\end{figure}

The distribution of relaxation times, $\tau$, is shown in Figure~\ref{tau}. For all sixty-two pages, $\tau$ is exceptionally long. The average relaxation time in our sample is $698$ steps, corresponding to a $\lambda_2$ of roughly $1-10^{-3}$; the median is $287$ steps. The effect size is large; the median relaxation time for the null model is only 12 steps, and for any particular page, the observed relaxation time is on average 50 times longer than the null expectation. In the majority of cases (52 of 62), there is significant evidence that global machine structure, in addition to transition sparseness, is leading to these long relaxation times. 

Using Viterbi reconstruction, we can infer the actual transitions that occur on the pages. Coarse-graining subspace transitions to remove flickering, we find a total of 1390 transition events over the sixty-two pages. The average trapping time, page-by-page, is 1201 steps or 190 days. \mbox{Trapping times} are longer than relaxation times. Epoch switching events in our sample do not appear to be concentrated at any special range of time in the last fifteen years; the distribution of switching times is indistinguishable, in a Kolmogorov--Smirnov test, from the distribution of edit times in general.

The two subspaces have distinct levels of expected reverts. We refer to the subspace with higher levels of reverting as ``high-conflict'' (type one) and the one with lower levels of reverting as ``low-conflict'' (type two). We emphasize that reverts track only one feature of conflict and that the lower conflict subspace not only shows significant amounts of reverting, but also that (as we shall see) the lower conflict subspace may have more examples of norm-violating conflict. We suggest use of the phrase ``type one conflict'', rather than ``conflict associated with being in the higher-conflict subspace'' (and similarly for ``type two conflict'') for simplicity. While we use this language informally, we refer the reader to our discussion above regarding the limitations of reverts as a tracer of more elaborate notions of conflict.

\begin{figure}[H]
\centering
\includegraphics[width=0.75\textwidth]{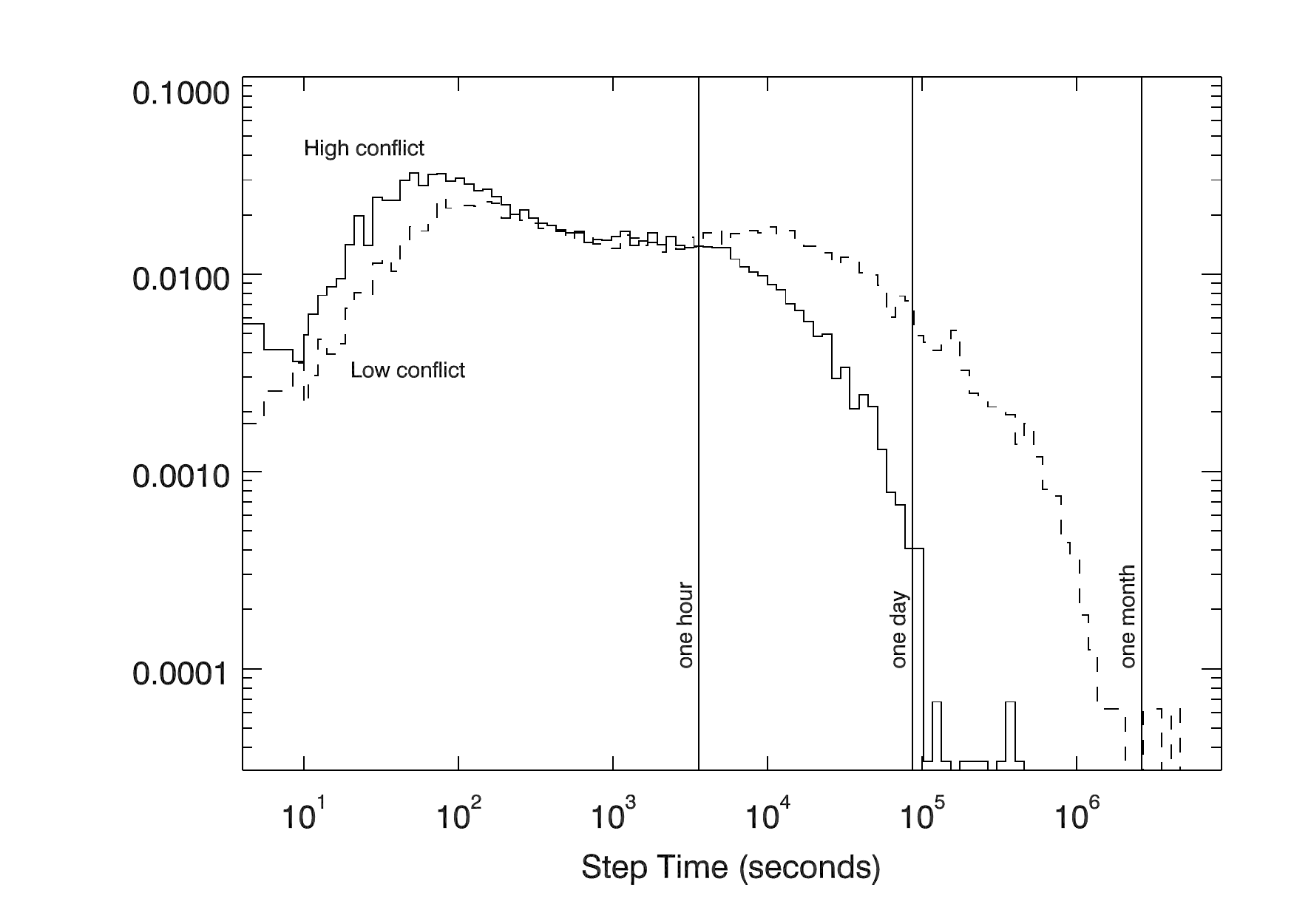}
%% because some colleagues have been confused about the meaning of "s", I have left the image with "seconds" rather than "s" for scientific reasons
\caption{Distribution of edit rates for type one (high conflict) and type two (low conflict) subspaces. Shown here, as an example, is the spacing between edits on the George W. Bush page. When the system is in the high-conflict subspace, edits occur once every 212 s (median; 3.5 min); in the low-conflict subspace, once every 951 s, or every 16 min. When the system is in the high-conflict subspace, users almost never wait more than a day to take action.\label{gwb_seconds}}
\end{figure}

On average, the higher-conflict subspace is roughly 2.5 times more revert-prone than the ``low-conflict'' subspace. Residency times are also slightly longer in the high-conflict state (207 days vs. 149 days). When the system is trapped in the higher conflict subspace, we find that editing usually, though not always, accelerates: the spacing between edits declines. In the mean, across all pages, users edit on average 2.6 times faster when in the higher conflict subspace; 46 of the 62 pages show this acceleration (high conflict faster than low conflict). The differences in edit rate are driven in part by a long tail of wait times; if we consider not the average time between edits, but the median time between edits, the difference shrinks, and only 37 of 62 pages show acceleration. An example of the distribution of edit timescales is shown in Figure~\ref{gwb_seconds}, for the case of the George W. Bush page; when the system is in the high-conflict subspace, edits occur almost every three minutes, compared to once every 16 min in the low-conflict subspace. Other differences appear; 37\% of edits are made by unregistered users (``IP addresses'') in the high-conflict subspace, compared to 21\% of edits in the low-conflict subspace.

As can be seen in the sample Viterbi reconstruction in Figure~\ref{gwb_picture}, the transition from one state to the other is characterized by more than just an increase in revert rate. High- and low-conflict subspaces are characterized by very different interaction styles. Table~\ref{motifs} shows the characteristic motifs over the full set of 62 pages. As expected, the low-conflict subspaces are distinguished by runs of cooperation; however, they are also, unexpectedly, distinguished by runs of reverts. When reverts follow each other, they are usually part of an edit war~\cite{yasseri2012dynamics}, involving a small number of users repeatedly undoing each other's work in a tit-for-tat pattern.

Conversely, vandalism (when a user inserts patently offensive or nonsense content into a page) followed by a (pro-social) revert is one of the characteristics of the high-conflict subspace; the vandal appears as a C move, and the pro-social repair as R. CR events are roughly 2.7 times as common when the system is in the high-conflict subspace, and edit comments that report vandalism (using the keywords ``vandal'' or the shorthand ``rvv'') are 1.9 times as common. %% these are defined

Not all CR motifs are simple vandalism, however. The underlying CR pattern is simply a rejected proposal; repeated CRs in a time series are a signature of repeated rejection of different proposals. As an example, we take a (randomly chosen) short period when the Hillary Clinton page is in the high-conflict subspace~\cite{clinton}; the system entered the high-conflict subspace on \mbox{8 February} 2007, 09:21 UTC, and left on 17 February 2007, 20:14 UTC). Of the eighty edits of that epoch, there are twelve CR pairs; six of them revert simple vandalism (introduction of sexual slurs), while five involve rejection of content that was contributed by a single editor apparently acting in good faith, and one is ambiguous. 

CR motifs are a significant source of the higher levels of system conflict: while 28\% of all edits in the high-conflict subspace are reverts, only 8\% of reverts remain once CR motifs are dropped. Removing CR motifs, however, increases the relative rates of conflict; the high-conflict subspace has 2.7 times more reverts than the low-conflict subspace when CR pairs are removed, compared to 2.5 in the base condition. When comparing conflict rates between the subspaces, the residue is worse than the mixture. 
\begin{table}[H]
\centering
\small
\begin{tabular}{cll}
\toprule
\textbf{Size} & \textbf{Type One Motifs} & \textbf{Type Two Motifs} \\ \midrule
2 & CR, RC & CC, RR \\
3 & CRC, RCR, RCC, CCR & CCC, RRR, RRC, CRR \\
4 & RCRC, CRCR, CRCC, CCRC, RCCR & CCCC, RRRR, RRCC, RCRR, CRRC \\
5 & {\small CRCRC, RCRCR, RCRCC, CCRCR, CRCCR} & {\small CCCCC, RRCRR, RRRRR, CRRRR, RRRRC} \\
\bottomrule
\end{tabular}
\caption{Characteristic motifs of the higher (type one) and lower (type two) conflict subspaces across all 62 pages, ranked by partial-KL (see Ref.~\cite{obo}, Eq. 2). The lower-conflict subspace is characterized by long runs of cooperation, but also by long runs of reversion. Conversely, the higher-conflict subspace is characterized by more rapid patterns of alternation between R and C moves. \label{motifs}}
\end{table}

Transitions are not accompanied by significant population shifts. On average, 20.2\% of the users that have appeared in the 100 edits just prior to a transition point also appear in the 100 edits just following. This is, in fact, slightly higher than the persistence we expect across an arbitrary point (17.7\%) in our data. Transitions to the high (or low) conflict subspace are not associated with unusual rates of turnover.

\subsection{Drivers of Conflict Transitions}

\textls[-15]{For the three event types that may be causal drivers of the transition to or from the high-conflict state---page protection, anti-social users, and external events---there are three potentially relevant questions:}

\begin{enumerate}
\item \label{effect} How many events are there and what fraction are associated with a transition? (Effectiveness) 
\item What fraction of transitions are associated with an event? (Explanatory power) \label{inf}
\item For those transitions that we can associate with an event, what fraction have the expected \mbox{effect?} (Valence) \label{logic}
\end{enumerate}

Item~\ref{effect} considers a measure of event effectiveness; an event type that more often leads to a transition is more effective. Item~\ref{inf} considers the extent to which an event type can explain an observed transition; if more transitions are associated with such an event type, that type is the more influential in explaining state switching. Item~\ref{logic}, valence, measures the extent to which the event has the expected effect; when an administrator locks a page, for example, it is natural to assume that the goal is to push the system out of a high-conflict state. Measures of valence only make sense when an event type is effective or has explanatory power.

These measures should be kept distinct, just as they are, for example, in the testing of a new drug. Taking a drug may correlate with a change in the patient's symptoms (apparent effect); a patient's symptoms may fluctuate independently of his or her use of the drug (low explanatory power); the drug may increase, rather than reduce, the severity of the symptoms (wrong valence).

\subsubsection{Page Protection Events}

Administrators can exercise direct control over who edits a page. These page protection events are the primary mechanism for authorities to influence the patterns of editing on the page; they are a blunt and top-down instrument that restricts editing on a particular page to increasingly smaller populations (``hard'' protection) or, conversely, opens it up to the wider community (``soft'' protection events). Releasing a page from protection can occur by explicit action or by default, when a prior restriction is given an expiration date. 

Page protection is mostly ineffective in inducing a transition. Of the 1545 protection events in our data, only 136 (8.8\%) occur within ten edits of a transition event; while the effect is statistically significant (35, or 2.3\% expected; $p<10^{-3}$), the effect size is small.

Page protection can explain some, but far from all, transition events. There are 1387 transition events that we can, potentially, associate with a page protection (this is less than the total number of transitions, because we do not have reliable records of protection events prior to 10 November 2003); thus, at most, about 9.8\% of observed transitions can be associated with page protections (2.5\%~expected; $p<10^{-3}$).

Norms on the encyclopedia itself urge the use of page protection as a response to conflict. \mbox{We thus} might expect protections to lead to a transition to the low-conflict state; and, conversely, weakening of protection to lead to a transition to the high-conflict state. This happens 80\% and 72\% of the time, respectively. Interventions are not usually successful in changing a page state, but when they do, their effects do have the correct valence. Page protection appears to operate in the direction \mbox{administrators expect}. 

Increasing the window increases the number of transitions we can potentially associate with a protection event. When we enlarge the window size to one hundred edits, 397 of the transition events, or 29\%, are now associated with a page protection event. The signal is statistically significant ($p<10^{-3}$), but the false detection rate is higher (240, or 17\% expected) and the valence less certain (65\% rather than 80\%). Post-selecting for the optimal window provides an upper bound: at most, only about 11\% of transitions can be reliably explained by page protection events.

Top-down control thus has at best only weak effects on the transition between the high- and low-conflict subspace. This is surprising, given the association between vandal-like CR patterns and the high-conflict subspace. ``Locking down'' a page may lighten the burden for pro-social users who police vandalism, but the high-conflict subspace describes more than just vandalism, and excluding users, even large classes of them, only rarely induces lasting effects on long-term patterns of cooperation and conflict. Conversely, opening up a page rarely leads to an emergence of new conflict. Most top-down actions are unsuccessful in causing a page transition; most transitions can not be explained by reference to top-down action. Comfortingly, however, when associations can be found, they do have the correct valence. It is unusual for a hard protection event, for example, to be associated with a transition to the high-conflict subspace.

\subsubsection{Anti-Social User Events}

A transition to the high-conflict state could be occasioned by the appearance of particularly anti-social users. Recall that anti-social users are defined by having an average blocking rate higher than 95\% of users who have appeared on the page. Anti-social users are said to ``dominate'' a transition when they make the plurality of edits within the window in question.

There are a total of 710 transitions to the high-conflict state in our data. Of those transitions, \mbox{212 are} associated, within 10 steps, with the dominance of an anti-social user, compared to an expected 208 in the null. We do not find above-null evidence of anti-social users triggering a transition ($p>0.1$). Anti-social user events do not explain the transitions we see in the data. Expanding the range to \mbox{100 steps} leaves the results unchanged: 255 transitions can be associated with an anti-social user event, compared to 263 in the null. Despite a significant focus on the management of problematic users within the encyclopedia~\cite{bradi, reagle2010good}, their appearance at a particular point does not appear to be a proximate cause of conflict.

This high null-expectation rate (35\%) implies that a significant fraction of edits, at least on high-traffic pages, is made by users who have received significant numbers of administratively-imposed sanctions compared to their total number of edits. The simple appearance of this kind of user, however, is ineffective in inducing a transition. They seem to play many roles in the system, including roles consistent with the system remaining in the low-conflict subspace, perhaps due to preference change and dynamical learning of social norms~\cite{major}.

Anti-social users do not explain transition events, and they are ineffective in inducing a transition.

\subsubsection{Major External Events}

Not all pages are associated with a topic sufficiently newsworthy that external events can be tracked; once we eliminate pages where less than 10\% of months have an article during the page lifetime, we are left with a sample of 59 pages.

Epoch transitions associated with external events include George W. Bush's election win in November 2004, Bob Dylan's first number one hit in thirty years in September 2006 and Barack Obama's election win in November 2008 and his inauguration in January 2009. However, many significant news events are not associated with epoch transitions, including the death of Michael Jackson in 2009 (which left other significant traces on the page~\cite{yasseri2012dynamics}) and John Kerry's election loss in November 2004. Our sample of most-edited pages on Wikipedia includes a page on the 2006 Lebanon War, begun during the war itself. Previous work on breaking news collaborations has found distinctive editing patterns associated with events of this form~\cite{keegan2012staying}; in the Lebanon case, the two transitions (to high conflict, on 9 October 2006; and back to low conflict, on 13 August 2013) occur after the war itself has concluded.

There are a total of 1367 transitions that we could potentially associate with newsworthy events. A total of 146 of these transitions are associated (10.6\%) with such an event. We expect, on average, 126 events in the null case (9.2\%). While there is weak evidence for some above-null association ($p\approx 0.03<0.05$), the effect size appears to be small: less than 2\% of all transition events can be reliably associated with outlier news coverage. 

Unusual spikes of real-world news coverage do not explain transition events, and, because of the lack of association, we can also assert that they are ineffective in inducing a transition.

\section{Discussion}

The results described here provide novel evidence for the existence of an epoch-like structure of conflict and cooperation on Wikipedia, distinguished by behavioral motifs. When we model page editing as a finite state machine, we find that pages can be trapped in one or another subset of their computational subspaces, often for hundreds of days at a time.

At the most coarse-grained level, this trapping places the system in either a low- or high-conflict state. In the high-conflict state, we also see an acceleration of activity by a factor of two or more. Subspaces are characterized by more than just overall levels of conflict: they have distinct motifs of interaction, including tit-for-tat (repeated reverting) in the low-conflict subspace, and propose-reject (C then R) and strong signals of vandal repair in the high-conflict subspace. 

The epoch structure revealed by our finite state analysis is defined by far more than just the density of reverts. As can be seen in the example time series in Figure~\ref{gwb_picture}, the low-conflict subspace can have long runs of reverting and edit warring. Low-conflict, in other words, does not mean zero conflict and may even mean norm violation; repeated reverting is likely to violate the ``three revert rule'' (3RR) that prohibits a single user from making three reverts to a page in a row. 

In game theoretic accounts of behavior on Wikipedia, reverts can be likened to ``defect'' moves, and repeated reverts by different users are something we might expect from a tit-for-tat like strategy under noise~\cite{group}. The association of long strings of reverts with the low-conflict subspace fits with accounts that put tit-for-tat at the heart of successful resolution of collective action problems from asocial beginnings~\cite{Axelrod27031981}. The fact that they violate the Wikipedia-specific 3RR norm suggests that this more ancient tit-for-tat strategy may still play an important functional role in managing conflict. Individuals may be banding together: past work on Wikipedia conflict has examined the influence of edit wars that draw in multiple mutually-reverting subgroups, and these more structured interactions may characterize the low-conflict subspace. In the case of edits to the Terri Schiavo page, for example, a network analysis finds three groups of antagonists: one group of administrators, and two groups associated with a particular point of view~\cite{kittur2007he}.

The patterns we describe here extend over thousands of edits; our epoch structure is on a longer timescale than the laws that describe short-term, repeated cooperation on timescales of minutes and hours~\cite{dedeo2013collective}. It is consistent with findings of the importance of long-term system memory in the bursty structure of edit wars~\cite{viegas2004studying,viegas2007,sumi2011characterization,sumi2011edit,yasseri2013value}. Our work goes beyond previous studies to explicitly construct an approximation to the full social grammar of system conflict~\cite{jackendoff2007language}. Since the number of users far exceeds the number of states, each state is an irreducible coarse-graining of an (implicit) fine-grained, complete account of the mental states of hundreds of individuals acting in a context set by the text on the page~itself. 

A natural language for the discussion of the transition between conflict and cooperation comes from the critical transitions literature. This work draws on tools from catastrophe theory, critical bifurcations, and phase transitions~\cite{scheffer2009early,lade2012early}. They focus on the unusual and potentially chaotic dynamics that emerge from the interaction of large numbers of heterogeneously-coupled units. These accounts are often mechanism-neutral and focus on abstract properties of a time series of events. In the literature on critical transitions, for example, one looks for properties such as spatial correlation~\cite{dakos2010spatial}, critical slowing down~\cite{dakos2008slowing} or flickering~\cite{wang2012flickering}, and can apply their logic to clinical depression just as well as ecosystem collapse~\cite{van2014critical}. In its focus on discrete states, rather than continuous fields, our work is complementary to this tradition. It can be thought of as the extension of linguistics to the construction of collectively-implemented social grammars~\cite{jackendoff2007language}.

We have focused on the binary classification revert/non-revert. Recent work~\cite{keegan2015analyzing} has studied the motifs associated with more fine-grained classifications of Wikipedia editing, focusing in particular on the nature of the user making the edit (for example: is this the user's first edit on Wikipedia, the first on this article, the first in this ``session''?). In a mixed strategy that combines quantitative and qualitative analysis, they looked at short timescale motifs of these more complex patterns associated with content co-production ``routines''~\cite{feldman2003reconceptualizing,pentland2015organizational}. Our work suggests that extending this analysis, using the hidden Markov model tools presented here, to longer timescales may well reveal correspondingly longer timescale epochs, which may or may not be nested within the conflict and cooperation structure we find here. Indeed, a parallel analysis of different time series would allow us to determine the extent to which different cultural practices coexist and interact: how the strategies people use to improve content vary in the presence of the strategies people use to defuse conflict. Indeed, one of the most influential essays in Wikipedia's norm network explicitly combines the two: the so-called ``BOLD, revert, discuss cycle''~\cite{bold}, which explicitly links content improvement and content reverting~\cite{bradi}.
%%is the capitalization necessary? please define if an acronym -- capitalization is necessary; it is not an acronym

Our work also connects to the now-extensive literature in organizational behavior on Wikipedia and Computer-Supported Collaborative Work (CSCW). Our findings on the dynamic switching between patterns of interaction is consistent with work that has emphasized the different ways that Wiki-like organizations can deal with resource flows that lead to either generative or constrained tensions~\cite{faraj2011}. Studies in this literature have emphasized the role of the community structure~\cite{kane2011,kanefichman2009} and describe systems like Wikipedia as a knowledge conversation~\cite{majchrak2013}, where the interaction between participants, and between participants and the wider context of content on the encyclopedia~\cite{sean,ransbothamkane2011,ransbotham2012}, rather than just between participant and content, determines the system's evolution over time.

We have chosen a particularly simple computational model for our system; formally, a system at the base of the Chomsky hierarchy~\cite{chomsky1965aspects}. A finite-state machine can, in short runs, approximate more sophisticated grammars, but the pumping lemma (and its simple probabilistic extension; see~\cite{5392601,dedeo2013collective}) means that, on sufficiently long timescales, it must fail. Indeed, our epoch-like structure already suggests the existence of long-range memory: repeated CR motifs, for example, in one month, correlate with the presence of CR motifs many months later. Examination of the detailed structure of our machines may suggest priors for higher-order grammars, such as the context-free languages that allow for nested pairing. 

Within the field of time series analysis itself, an open question concerns the relationship between these simple finite-state models and the epsilon machines~\cite{crutchfield1989inferring,crutchfield1994calculi,crutchfield1999thermodynamic}; while hidden Markov models generically imply infinite-state epsilon machines, recent progress has suggested ways to lossily compress these representations to a finite system~\cite{marzen2014circumventing}. These results currently suggest that, because of the exceedingly long relaxation times (small spectral gaps) seen in Wikipedia, estimating some of the quantities relevant to the computational mechanics paradigm is impossible even for the very longest sequences ($\sim$$10^4$ data points) we can observe. However, other recent work in this tradition suggests that, with a sufficiently explicit phenomenological model, estimation of quantities such as excess entropy may be possible~\cite{marzen2}. Our current belief is that the discovery of these very small spectral gaps suggests that a finite-state machine is only an approximate model, and that the true nature of memory storage and processing is likely to be more interesting yet. Its description may demand novel mathematical structures beyond the simple cases currently found in the quantitative literature.

We examined three {prima facie} plausible mechanisms for switching the system between these two states: top-down administrative actions, the appearance of unusually anti-social editors and major external events in the real world. In only one case, administrator action, did we find statistically-significant associations compatible with a causal role for state-space switching. 

One of the characteristics of the high-conflict subspace are motifs often (though not always) associated with vandalism---the repeated CR pairs. Part of the explanation of the transition to the high-conflict subspace ought to involve new sources of this particular behavior. Efforts to exclude anonymous IP addresses, or registered but untrusted users, suggest that administrators are able to recognize sources of vandalism and intervene to prevent them. However, while vandalism is more commonly associated with the higher-conflict type one state, the majority of edits do not show signatures of vandalism, and locking down a page is largely ineffective in shifting the system to the lower-conflict type two state. Moreover, transitions in (or out) of the higher conflict state are not associated with higher user turnover and so cannot be explained by an unexpected influx of \mbox{new users}. 

Top-down effects are not the only thing that can drive pages between higher and lower conflict states: the effects of multiple user-user and user-page interactions can conspire to leave long-term traces on the page, as shown by the mathematical model of~\cite{torok2013opinions,iniguez2014modeling}. In this model, user behavior is dictated by a set of parameters that describe when a user will attempt to alter the page or, conversely, adapt to the page as it stands and, via discussion, to the opinions of others. When the user population is fixed, this process drives the system to consensus. However, when user turnover is sufficiently high (and the tolerance of different opinions sufficiently low), the system can be driven into a state of permanent conflict. Between these two phases is a line of critical coexistence, where, paralleling what we see in our analysis here, periods of consensus are interrupted by stretches of war.

These simple mathematical models provide an intriguing way for explaining our results. \mbox{The bistability} we observe empirically may be the consequence of an underlying process of belief revision when augmented by the presence of a common resource for information sharing. The question then becomes: what drives Wikipedia to the critical line separating these two phases?

While we are able to rule out spikes in news coverage as deterministic exogenous drivers, one of the key limitations of our work here is our inability to rule out more complicated interactions with the larger world. Long-term system memory may reside in users, in the text of the page, or well beyond the bounds of the encyclopedia. In the end, Wikipedia is not an isolated system. Its internal logic is coupled to the wider world, one rich in its own bursty and autocorrelated structure~\cite{b1, b2, b21, b3, b4}, and there is no clear division between the two~\cite{wellman96,Wellman2031}. It is this joint process we observe and must analyze: \mbox{a small} and shifting fragment of the Internet as a whole.

\section{Conclusion}

Conflict is endemic to social life. Rather than exclude conflict altogether, however, most societies attempt to manage the ways in which conflict and cooperation interact. The co-existence of conflict and cooperation is a basic theme of studies of intra-group conflict in the biological sciences~\cite{flack2005robustness,flack2006policing}. Quarreling and fighting are not simply unstructured forms of letting off steam. They are complex phenomena that make significant cognitive demands on individuals~\cite{dedeo2010inductive,hobson} and that result in epochs that decouple from day-night cycles and far outlast any particular clash~\cite{dedeo2011evidence,flack2012multiple}. Wikipedia and IT-mediated social participation systems more generally, are unusual in the extent to which their underlying system norms foreground, and even valorize, conflict during co-production. In their broader structures, however, the phenomena we see here would be familiar to researchers studying regularities in conflict across the biological world.

\textls[-10]{Finite-state modeling of Wikipedia's social grammar provides new insight into how individual-level} complexity leaves a trace on the system as a whole. In a large-scale, virtual environment, with many tens of thousands of participants drawn from the wider Internet, we find that conflict retains a complex logical structure, with context-dependent system memory, which means that any particular event must be understood in a wider context that extends many months back, and may help define the system's future years hence. The dynamics of this memory and its associated patterns of conflict cannot be captured by simple top-down accounts.

Our results indicate new points of contact for the mathematical, social, and biological sciences. By measuring the essentially computational properties of a social world---its spectral gap or the logic of its subspaces---we gain new insight into the ways in which our species can trap itself in the very patterns it creates.

\acknowledgments{I thank Bradi Heaberlin (Indiana University), Taha Yasseri (Oxford Internet Institute), \mbox{Brian Keegan} (Harvard Business School), Cosma Shalizi (Carnegie Mellon University), and my three anonymous referees for comments on early versions of this manuscript, as well as the Global Brain Institute of Vrije Universiteit Brussel, Belgium, at which this work was presented. I thank Nate Metheny (Santa Fe Institute) for the construction of the machines Ganesha, Saraswati and Laxmi, which made the calculations in this paper possible. This work was supported in part by National Science Foundation Grant EF-1137929.}

\conflictofinterests{The author declares no conflict of interest.}

\appendixtitles{yes}
\appendix
\counterwithin{figure}{section}
\counterwithin{table}{section}
\renewcommand{\theequation}{\Alph{section}-\arabic{equation}}

\section{Articles in Our Analysis}
\label{full_list}

The number of edits in each page's time series is listed in parentheses after the article title, along with a simple classification: George\_W.\_Bush (45,448; biography, politician), United\_States (33,725; geography), Wikipedia (32,592; technology), Michael\_Jackson (27,587; biography, entertainment), Catholic\_Church (24,813, religion), Barack\_Obama (23,889; biography, politician), World\_War\_II (23,173; event, political), Global\_warming (20,003; science), 2006\_Lebanon\_War (19,972; event, political), Islam (18,523; religion), Canada (18,150; geography), Eminem (18,066; biography, entertainment), September\_11\_attacks (17,564; event), Paul\_McCartney (16,973; biography, entertainment), Israel (16,790; geographic), Hurricane\_Katrina (16,753; event), Xbox\_360 (16,753; technology), Pink\_Floyd (16,037; biography, entertainment), Iraq\_War (15,891; event), Blackout\_(Britney\_Spears\_album) (15,832; entertainment), Turkey (15,663; geography), Super\_Smash\_Bros.\_Brawl (15,432; technology), World\_War\_I (15,292; event), Gaza\_War (14,920; event), Lost\_(TV\_series) (14,897; entertainment), Blink-182 (14,789; entertainment), Scientology (14,727; religion), John\_Kerry (14,307; biography, political), Heroes\_(TV\_series) (14,223; entertainment), Australia (14,186; geography), China (14,023; geography), Bob\_Dylan (13,916; biography, entertainment), Neighbors (13,547; entertainment), The\_Holocaust (13,346; event), Atheism (13,295; religion), Hilary\_Duff (13,222; biography, entertainment), Mexico (13,213; geography), The\_Dark\_Knight\_(film) (13,025; entertainment), France (12,800; geographic), John\_F.\_Kennedy (12,788; biography, politician), Lindsay\_Lohan (12,757; biography, entertainment), Girls'\_Generation (12,746; entertainment), Argentina (12,745; geography), Virginia\_Tech\_massacre (12,682; event), RMS\_Titanic (12,451; event), Russo-Georgian\_War (12,365; event), Homosexuality (12,170; science), Circumcision (12,149; religion, science), Hillary\_Rodham\_Clinton (11,981; biography, politician), Star\_Trek (11,919; entertainment), Shakira (11,712; biography, entertainment), Sweden (11,666; geography), New\_Zealand (11,639; geography), Paris\_Hilton (11,635; biography, entertainment), Wizards\_of\_Waverly\_Place (11,520; entertainment), Genghis\_Khan (11,410; biography, politician), Cuba (11,390; geography), Linux (11,316; technology), Che\_Guevara (11,250; biography, politician), Golf (11,141; entertainment), iPhone (11,085; technology), God (10,731; religion).

\section{Choosing the Number of States in an HMM}
\label{model_selection}

\vspace{-6pt}
\begin{table}[H]
\centering
\small
\begin{tabular}{ccccccccccc} 
\toprule
\textbf{Number of States} & \textbf{1} & \textbf{2} & \textbf{3} &\textbf{ 4} & \textbf{5} & \textbf{6} & \textbf{7} & {\bf 8} (Truth) & \textbf{9} & \textbf{10} \\ \midrule
AIC & 0.0\% & 0.0\% & 0.0\% & 0.0\% & 0.0\% & 37.5\% & 7.2\% & 45.8\% & 9.3\% & 0.0\% \\ 
BIC & 0.0\% & 0.0\% & 0.0\% & 20.8\% & 54.1\% & 23.9\% & 1.0\% & 0.0\% & 0.0\% & 0.0\% \\ 
\bottomrule
\end{tabular}
\caption{Using AIC and BIC to choose the number of states in a hidden Markov model fit. Here, we take an actual model from our data (the 8-state best fit model for the {\tt God} page), use that model to generate a new time series of equal length (10,731 samples) and attempt to fit a new model, using either AIC or BIC to select the preferred number of states in a manner similar to Refs.~\cite{celeux2008selecting,bacci2014comparison}. The table lists the fraction of the time this process led to a preferred machine of each size, for the two different penalties. Both AIC and BIC tend to underestimate model complexity; in general, BIC performs worse, significantly underestimating the true number of states. AIC performs better, recovering the correct number of states nearly half the time.} %% this table is original to this paper; therefore, no copyright is involved
\end{table}

Choosing the number of states to include in an HMM is an example of a model selection problem; one is selecting between models with different numbers of states. In general, the larger the HMM, the better the data can be fit: when does one stop improving the fit because the model is becoming ``too complex''? A generic solution to model selection is cross-validation: one fits the model using a subset of the data (the ``training set''), and sees how well the model performs when predicting out of sample (the ``test set'' or ``hold-out'' set).

Cross-validation uses the phenomenon of over-fitting to determine when a model is too complex: if a model has too many parameters, it will overfit to the training set, finding ``patterns'' that are really due to coincidence. These patterns will fail to hold in the test set, and will degrade performance on the fit; this degradation can be measured, and one stops increasing model complexity when performance on the test set first starts to decline. Standard cross-validation techniques work best when the data are independently sampled, i.e., when it is possible to construct a test set that is uncorrelated with the training set conditional on the underlying model. As increasing levels of correlation appear in the data, the construction of an appropriate training set becomes difficult.

When, as is the case for HMMs, the model is Bayesian, another method is possible: the likelihood penalty. To use a likelihood penalty, you fit the model to all of the data and note the posterior log-likelihood. You then apply a penalty, reducing the log-likelihood depending on features related to the complexity of the model, including, usually, the total number of parameters. After applying this penalty, it is usually the case that one particular model, often not the most complex one, maximizes the penalized log-likelihood, and this is the one considered preferred. Numerous likelihood penalties exist, including the Bayesian Information Criterion (BIC~\cite{schwarz1978estimating}), the Bayesian Evidence (introduced in~\cite{mackay2003information}; used on Wikipedia data in~\cite{dedeo2013collective, group}) and the Akaike Information Criterion (AIC; introduced in~\cite{akaike1974new}).

The existence of both long- and short-range correlations in data fit using an HMM makes the use of cross-validation and hold-out techniques difficult. The work in~\cite{celeux2008selecting} proposes 
two methods for cross-validation on HMMs that attempt to compensate for the failure of independence. Both methods have difficulties, and work well only on a subset of HMMs, where correlations are not particularly long-range and the transition matrix is not too sparse. Because we find that both conditions are violated in our data, we do not attempt cross-validation tests. The work in~\cite{celeux2008selecting} also considers the AIC and BIC methods; they find that both work well in recovering the true size of an HMM used to generate simulated data. We are not aware of work that has tested the use of Bayesian Evidence on simulated data and defer this interesting, but involved, question to later work.

The authors of Ref.~\cite{celeux2008selecting} find that, when it fails, BIC tends to slightly underestimate, and AIC to slightly overestimate, the true number of states, although both criteria work well. Work by~\cite{bacci2014comparison} confirms this result. However, both papers consider regimes that do not directly apply here, and both papers found cases where (for example) BIC significantly underestimated model complexity. \mbox{In addition}, both papers consider problems with an order of magnitude less data than we have and true model HMM sizes much smaller than ten. In the main text, we use the AIC penalty, a common choice made across the biological and signal processing communities, and strongly argued for on general grounds by~\cite{burnham2003model} when the goal is minimizing prediction loss. 

To validate our choice and, in particular, to determine whether AIC or BIC produces more valid results for the regimes relevant here, we did an in-depth test with a particular model: the eight-state HMM associated with the {\tt God} page. We took the derived HMM for this page and used it to generate 96 simulated datasets of equal length to the original. We then ran our fitting code on each of these datasets and compared what happened when we used the AIC and the BIC criteria to select the preferred number of states. Consistent with~\cite{celeux2008selecting,bacci2014comparison}, we found that BIC tended to underestimate model complexity, choosing machines significantly smaller than the true number and, in fact, never recovering the true system size. We found, by contrast, that AIC worked better; like BIC, it still often underestimated model complexity, but did so by smaller amounts. Conversely, a small fraction of the time (less than 10\%), it preferred a model that was one state more complex than reality.

Because our main concern is to minimize overfitting and the introduction of fictitious structure, without losing too much of the actual structure in the process, AIC's slight tendency to underestimate model complexity is not a major concern. A small fraction of the time (roughly 10\%), AIC preferred a machine that was one state larger; however, robustness tests show that none of our main conclusions depend on the exact number of states being that chosen by the AIC criterion. We recover the same conclusions, for example, if we arbitrarily fix the number of states equal to twelve. It is worth noting that we do not believe the ``true'' model of Wikipedian conflict is itself a finite-state machine~\cite{dedeo2013collective}; i.e., the fundamental problem is finding the best approximation, rather than locating the correct model within a known class.

\section{Relaxation Time, Mixing Time, Decay Time, Trapping Time}
\label{formal_lambda2}
\setcounter{equation}{0}
By the Levin--Peres--Wilmer theorem~\cite{levin2009markov}, the relaxation time, $\tau$, defined in Equation~(\ref{relax}) above, provides an upper bound on the mixing time, $\tau^\prime(\epsilon)$, the maximum time it takes an arbitrary initial condition to be no further than a small distance $\epsilon$ from the stationary distribution, where distance is defined as the maximum absolute value difference for any state occupation probability. In particular, we have
\begin{equation}
(\tau-1)\log\left(\frac{1}{2\epsilon}\right) \leq \tau^\prime(\epsilon) \leq \tau\log\left(\frac{1}{\epsilon\pi_\mathrm{min}}\right),
\end{equation}
where $\pi_\mathrm{min}$ is the smallest value in the stationary probability distribution; of characteristic order $10^{-1}$ in the HMMs considered in this paper. These relationships hold when the HMM is ``reversible'' and ``irreducible''; in empirical work, such as that presented here, those conditions are almost always satisfied. For example, they hold when states in an HMM have a small self-loop probability, no matter how small, and a directed path exists between any two states; both of which are true by default when using a Dirichlet prior and are true by inspection in our EM fits. This justifies a useful and intuitive interpretation of $\tau$ as a measure of how quickly an arbitrary initial condition converges to the average, as well as suggesting more sophisticated measures that take into account the allowable level of deviation ($\epsilon$) and inhomogeneities in the stationary distribution itself ($\pi_\mathrm{min}$).

The relaxation time $\tau$ is related to another natural quantity, the decay time of the second eigenvector. The time constant for decay, $\tau_d$, is related to $\lambda_2$ as
\begin{equation}
\lambda_2^t = e^{-t/\tau_d},
\end{equation}
which implies that
\begin{equation}
\tau_d = -\frac{1}{\ln\lambda_2}.
\end{equation}

On doing a Laurent series expansion around $\lambda_2$ equal to unity, we find
\begin{equation}
\tau_d = \frac{1}{1-\lambda_2} -\frac{1}{2} + \mathcal{O}(\lambda_2-1) +\ldots,
\end{equation}
which implies that, to zeroth order in $\lambda_2-1$ (i.e., when relaxation times are long), $\tau_d\approx\tau-1/2$ as used in Equation~(\ref{decay}). All of the machines in this paper are in the regime where the relaxation time and decay time are nearly equivalent.

In contrast to relaxation, mixing, and decay time, trapping time is an empirically-measured quantity. To compute trapping time, you define a set of internal states of interest, and then use the Viterbi algorithm to reconstruct the maximum-likelihood path through a particular time series. Trapping time is then defined as the average length of time the system spends in the set of interest before it leaves. In this paper, we track trapping time for the two main subspaces as defined by the sign structure of the second eigenvector.

\bibliographystyle{mdpi}
\renewcommand\bibname{References}

\end{document}